\documentclass[a4paper,12pt]{article}

\usepackage{graphicx}
\usepackage{amsmath}
\usepackage{amssymb}
\usepackage{cite}
\usepackage{multirow}
\usepackage{float} 
\usepackage[british]{babel}
\usepackage{microtype}
\usepackage[utf8x]{inputenc}

\begin{document}

\vspace*{1.0truecm}
\begin{center}

{\large\bf Dark matter in the NMSSM with small $\lambda$ and $\kappa$}\\
\
\\
{\large M. M. Almarashi$^{a}$, F. Alhazmi$^{a}$, R. Abdulhafidh$^{a}$, S. Abdul Basir$^{b}$}\\
\vspace*{0.5truecm}
$^{a}$Department of Physics, Faculty of Science,  Taibah University, P.O.Box 344 , Medinah, KSA
\\
 $^{b}$Department of Computer Science, Faculty of Computer Science and Engineering, 
 Taibah University, P.O.Box 344 , Medinah, KSA
 \end{center}

\date{\today}

\begin{center}
\begin{abstract}
The NMSSM provides an excellent dark matter candidate, usually the lightest neutralino ($\tilde{\chi}^0_1$) being the lightest supersymmetric particle (LSP), in the universe.
It is a mixture of the bino, the neutral wino, the neutral higgsinos, and the singlino.
In this work, we investigate the features of the lightest neutralino by concentrating on a specific region of the NMSSM
parameter space with small values of both $\lambda$ and $\kappa$, taking into account theoretical and experimental constraints, including
relic density, direct and indirect detection constraints.  
We have found that the LSP dark matter is a singlino-dominated neutralino for most of the sample points
of the chosen scenario. Thus, contrary to expectations, the NMSSM  is not comparable to
the MSSM in terms of dark matter properties as $\lambda$ and $\kappa$ are very small. The LSP can also be either higgsino-dominated or
bino-dominated in some parameter space. The most important parameters affecting the mass and properties of the LSP 
are $\kappa$, $\lambda/\kappa$, and $\mu_{\rm eff}$. Further, we notice
that $\sigma^{SI}_p \approx  1.02$ $\sigma^{SI}_n$ and $\sigma^{SD}_p \approx  0.76$ $\sigma^{SD}_n$ for all the surviving sample points. 

\end{abstract}
\end{center}

\section{Introduction}
\label{sect:intro}
The discovery of the standard model (SM)-like Higgs boson with a mass around 125 GeV by the ATLAS \cite{ATLAS1-Higgs,ATLAS2-Higgs} 
and CMS \cite{CMS1-Higgs,CMS2-Higgs} experiments confirmed that the Higgs mechanism is the origin of the masses of subatomic particles. Most of the SM predictions
have been confirmed experimentally with high accuracy.
Despite its great success, the SM is  imperfect. It
has been regarded as a theory of only low energy effectiveness, and it contains many flaws.
One of its major flaws is the so-called hierarchy problem: the vast gap between the electroweak scale and the Planck scale. 
Moreover, the SM does not include the gravitational
interactions nor why it is considerably weaker than the other interactions. Hence, gravity requires further explanation. Additionally, the SM predicts neutrinos to be massless,
despite recent experiments proving that neutrinos oscillate and therefore inevitably have mass \cite{Super-Kamiokande:1998qwk}.
Further, the SM does not explain astronomical observations of the dark matter (DM) that makes up roughly $27\%$ of the mass-energy
budget of the universe \cite{Planck:2015fie,Planck:2018vyg}. 
Accordingly, it is expected that a new physics beyond the SM is present at high energy scales for various
theoretical and experimental reasons.

In order to address the shortcomings of the SM, several models of particle physics beyond the standard model (BSM) have been developed. 
One of the most appealing candidates is supersymmetric (SUSY) extensions of the SM, for reviews see: e.g., \cite{Nilles:1983ge,Haber:1984rc,SUSYreview},
which can alleviate the hierarchy problem by introducing contributions from supersymmetric particles to the Higgs mass term.
Further, in R-parity conserving supersymmetric models such as
the minimal supersymmetric standard model (MSSM), for reviews, see: e.g., \cite{MSSMreview}, the lightest supersymmetric particle (LSP), usually the lightest
neutralino ${\tilde{\chi}^0_1}$, is a stable weakly interacting massive particle (WIMP), neutral and cold, i.e., non-relativistic during galaxy formation, and accordingly acts as 
a good candidate for dark matter. The MSSM is probably one of the most popular BSM scenarios, yet it suffers from two critical problems:
the $\mu$-problem in the superpotential \cite{Kim:1983dt} and the little hierarchy problem \cite{Bastero-Gil:2000gmf}.

The simplest supersymmetric extension of the SM beyond the MSSM that can solve the above two critical problems of the MSSM is the
 next-to-minimal supersymmetric standard model (NMSSM)
\cite{NMSSM1,NMSSM2,NMSSM3,NMSSM4,NMSSM5,NMSSM6,NMSSM7,NMSSM8,NMSSM9,upper,NMSSM10,NMSSM11}. This scenario includes a Higgs singlet
chiral superfield $\tilde S$ in addition to the usual two MSSM-type Higgs
doublets $\tilde H_u$ and $\tilde H_d$, giving rise to 
seven Higgs states and five neutralinos compared to only five Higgses and four neutralinos
in the MSSM. When the scalar component of the singlet superfield develops a vacuum expectation value (VEV), an `effective' $\mu$-term, $\mu_{\rm eff}$, is
automatically generated with the magnitude of electroweak scale \cite{NMSSM4}.
Furthermore, the little hierarchy problem in the NMSSM can be relieved 
\cite{Bastero-Gil:2000gmf,Dermisek:2005ar} since the SM-like Higgs squared mass at tree
level is enhanced by an additional term proportional to the singlet-doublet coupling $\lambda$.

In the NMSSM, due to the introduction of the Higgs singlet superfield ($\hat S$), which leads to an additional fermionic
partner of $\hat S$ (singlino), the neutralino sector of the NMSSM is phenomenologically richer than that of the MSSM.
The larger diversity of the neutralino sector of the NMSSM 
has led to many studies in this direction , see e.g. \cite{Abel:1992ts,Stephan:1997rv,Stephan:1997ds,Cerdeno:2004xw,Belanger:2005kh,Gunion:2005rw,Ferrer:2006hy,
Cerdeno:2007sn,Hugonie:2007vd,Belanger:2008nt,Cerdeno:2009dv,Das:2010ww,Kang:2010mh,Cao:2011re,Cerdeno:2011qv,Carena:2011jy,
AlbornozVasquez:2012px,Das:2012ys,Kozaczuk:2013spa,Badziak:2015exr}.
In the NMSSM  with a $Z_3$ symmetry, the ${\tilde{\chi}^0_1}$ is most likely to be higgsino-dominated or singlino-dominated dark matter
\cite{Cao:2016nix,Ellwanger:2016sur,Xiang:2016ndq,Baum:2017enm,Ellwanger:2018zxt,Domingo:2018ykx,Baum:2019uzg,Abdallah:2019znp,Cao:2019qng,Guchait:2020wqn,
Barman:2020vzm,Cao:2021ljw,Zhou:2021pit}.
However, dedicated studies on a possible bino-dominated dark matter in the context of the NMSSM are lacking \cite{Abdallah:2020yag}.
On the experimental side, given the constraints of the most recent experiments from dark matter searches from the LHC
 \cite{CMS:2018xqw,CMS:2018szt,CMS:2017moi,ATLAS:2018qmw,ATLAS:2018eui,ATLAS:2019wgx},
 the WMAP/Planck experiments \cite{Planck:2015fie,Planck:2018vyg} and the dark matter direct detection and
indirect detection experiments \cite{LUX:2017ree,XENON:2018voc,XENON:2019rxp,PICO:2019vsc,Fermi-LAT:2015att}, large regions of the NMSSM parameter space
have been strongly constrained.

The purpose of this work is to investigate the features of the lightest neutralino as a candidate for the dark matter
by concentrating on a specific region of the NMSSM parameter space with small values of both
the singlet-doublet mixing coupling $\lambda$ and the singlet cubic interaction coupling $\kappa$, 
taking into consideration theoretical
and experimental constraints. For this scenario with small $\lambda$ and $\kappa$, the singlino component of the LSP is small so the NMSSM is MSSM-like.

In our scenario, we classify our results into three types: for type I, the lightest neutralino as the dark matter candidate
is a singlino-like, for type II, it is a higgsino-like, and for type III, it is a bino-like. We then determine the properties
of the parameter space for each type. Such a study would be helpful to understand the physics of dark matter.
The previous phenomenological studies are concentrated on the assumption that the lightest neutralino is
either a singlino-dominated or a higgsino-dominated dark matter with only a few studies mentioned to a bino-dominated dark matter,
while generally scanning the NMSSM parameter over a wide range of $\lambda$ and $\kappa$. Hence, the wide-ranging scanning results
could possibly lead to a limited understanding of the dark matter properties in our specific scenario with small $\lambda$ and $\kappa$.

The paper is organized as follows. In the next section, we give a brief description of the Higgs and neutralino sector of the NMSSM.
In section 3, we provide the characteristics of the parameter space of interest.
The numerical results are presented and discussed in section 4 to show the characteristic features of dark matter physics. Finally, we conclude in section 5.

\section{ A brief review of the Higgs and neutralino sectors of the NMSSM}

It is previously stated that the NMSSM contains one extra singlet superfield $\hat S$ besides the usual two Higgs doublets of the MSSM ($\hat H_u$ and $\hat H_d$).
Consequently, the associated superpotential of the NMSSM in its simplest form is given by
 \begin{equation}  
 W_{\rm NMSSM} =  W_{\rm MSSM} + \lambda\hat S\hat H_u\hat H_d 
+\frac{1}{3}\kappa{\hat S}^3,
\end{equation}
where $ W_{\rm MSSM}$ is the MSSM superpotential. The term $\lambda\hat S\hat H_u\hat H_d$ is introduced to solve the $\mu$-problem of the MSSM superpotential while the term $\frac{1}{3}\kappa{\hat S}^3$ is added to
avoid the Peccei-Quinn symmetry \cite{Peccei:1977hh, Peccei:1977ur}.

The soft supersymmetry (SUSY) breaking terms for both the doublet and singlet fields are given by
\begin{equation}
V_{\rm NMSSM}=m_{H_u}^2|H_u|^2+m_{H_d}^2|H_d|^2+m_{S}^2|S|^2
             +\left(\lambda A_\lambda S H_u H_d + \frac{1}{3}\kappa A_\kappa S^3 + {\rm h.c.}\right),
\end{equation}
where $A_\lambda$ and $A_\kappa$ are the soft SUSY breaking trilinear coupling parameters of the order of SUSY mass scale $m_{\rm{SUSY}}$.

After the electroweak symmetry breaking, the Higgs fields acquire non-zero vacuum expectation values (VEVs) 
$\langle H_u\rangle =\frac{1}{\sqrt{2}} \upsilon_u$, 
 $\langle H_d\rangle = \frac{1}{\sqrt{2}} \upsilon_d$ and $\langle S\rangle = \frac{1}{\sqrt{2}}\upsilon_s$.
As a result, the potential has terms for the non-zero mass modes for the scalar Higgs fields $S_i (i = 1, 2, 3)$,
pseudoscalar Higgs fields $P_i (i = 1, 2)$ and charged Higgs fields $h^\pm$, and can be given by
\begin{equation}
V_{mass}= \frac{1}{2}(S_1\ \   S_2\ \ S_3)\mathcal{M}_S\left( \begin{array}{ccc}S_1 \\ S_2 \\ S_3 \end{array} \right)
 +\frac{1}{2}(P_1\ \   P_2)\mathcal{M}_P\left( \begin{array}{ccc}P_1 \\ P_2 \end{array} \right)
 + m^2_{h^\pm}h^+h^-.
\end{equation}
One can obtain the tree-level masses of physical mass eigenstates as follows. The mass matrix for the neutral scalar
Higgs states reads \cite{MNZ}
\begin{equation}
 \mathcal{M}^2_{S11}=m^2_A+\bigg (m^2_Z-\frac{1}{2}(\lambda\upsilon)^2\bigg ){\rm sin}^22\beta,
\end{equation}
\begin{equation}
 \mathcal{M}^2_{S12}=-\frac{1}{2}\bigg (m^2_Z-\frac{1}{2}(\lambda\upsilon)^2\bigg ){\rm sin}4\beta,
\end{equation}
\begin{equation}
\mathcal{M}^2_{S13}=-\frac{1}{2}\bigg (m^2_A{\rm sin}2\beta+2\frac{\kappa{\mu^2_{\rm eff}}}{\lambda}\bigg )\bigg (\frac{\lambda\upsilon}{\sqrt{2}\mu_{\rm eff}}\bigg ){\rm cos}2\beta 
\end{equation}
\begin{equation}
 \mathcal{M}^2_{S22}=m^2_Z{\rm cos}^22\beta+\frac{1}{2}(\lambda\upsilon)^2{\rm sin}^22\beta,
\end{equation}
\begin{equation}
 \mathcal{M}^2_{S23}=\frac{1}{2}\bigg (4{\mu^2_{\rm eff}}-m^2_A{\rm sin}^22\beta-\frac{2\kappa{\mu^2_{\rm eff}}{\rm sin}2\beta}{\lambda}\bigg )\frac{\lambda\upsilon}{\sqrt{2}\mu_{\rm eff}}
\end{equation}
\begin{equation}
\mathcal{M}^2_{S33}=\frac{1}{8}m^2_A{\rm sin}^22\beta\frac{\lambda^2\upsilon^2}{\mu^2_{\rm eff}}+4\frac{\kappa^2{\mu^2_{\rm eff}}}{\lambda^2}
+\frac{\kappa A_\kappa \mu_{\rm eff}}{\lambda}-\frac{1}{4}\lambda\kappa\upsilon^2{\rm sin}2\beta,
\end{equation}
where 
 $m^2_A=\sqrt{2}\frac{\mu_{\rm eff}}{\sin2\beta}\bigg(A_{\lambda}+\frac{\kappa\mu_{\rm eff}}{\lambda}\bigg)$,
tan$\beta=\frac{\upsilon_u}{\upsilon_d}$ and $\upsilon^2 = {\upsilon^2_u} + {\upsilon^2_d} $.

Similarly, the mass matrix for the neutral pseudo-scalar Higgs states reads \cite{MNZ}
\begin{equation}
\mathcal{M}^2_{P11}=m^2_A,
\end{equation}
\begin{equation}
 \mathcal{M}^2_{P12}=\frac{1}{2}\bigg (m^2_A{\rm sin}2\beta-6\frac{\kappa{\mu^2_{\rm eff}}}{\lambda}\bigg )\frac{\lambda\upsilon}{\sqrt{2}\mu_{\rm eff}},
\end{equation}
\begin{equation}
\mathcal{M}^2_{P22}=\frac{1}{8}\bigg (m^2_A{\rm sin}2\beta+6\frac{\kappa{\mu^2_{\rm eff}}}{\lambda}\bigg )\frac{\lambda^2\upsilon^2}{\mu^2_{\rm eff}}{\rm sin}2\beta
-3\frac{\kappa\mu_{\rm eff}A_\kappa}{\lambda}.
\end{equation}

Finally, the mass of charged Higgs fields at tree-level is given by \cite{MNZ}
\begin{equation}
 m^2_{h^\pm}=m^2_A+m^2_W-\frac{1}{2}(\lambda \upsilon)^2.
\end{equation}

 In the MSSM limit, one can obtain the pure singlet states with the following masses
 \begin{equation}
    \mathcal{M}^2_{S} = 4\frac{\kappa^2{\mu^2_{\rm eff}}}{\lambda^2}
+\frac{\kappa A_\kappa \mu_{\rm eff}}{\lambda}, 
 \end{equation}
 
\begin{equation}
 \mathcal{M}^2_{P} = -3\frac{\kappa\mu_{\rm eff}A_\kappa}{\lambda}.
\end{equation}

It is clear from the above equations that at tree level the NMSSM Higgs sector is described by the six independent parameters:
$\lambda$, $\kappa$, tan$\beta$, $\mu_{\rm eff}$ , $A_\lambda$ and 
$A_\kappa$. Assuming  the CP-conserving case, the mass of the MSSM-like light Higgs boson at tree level is bounded by \cite{NMSSM4,NMSSM5}
\begin{equation}
 m^2_{h} < m^2_Z {\rm cos}^2(2\beta)+\frac{\lambda^2 \upsilon^2}{2}{\rm sin}^2(2\beta).
\end{equation}
The effect of the last term in this equation is to enhance the mass of the SM-like Higgs boson with up to 15 GeV higher than the corresponding one of the MSSM.
Clearly, low values of tan$\beta$ and large values of $\lambda$ are preferred to obtain a large value of the $h$ mass at tree level.
The scenario with $m_{h} < 125$ GeV means that the $h_1$ is highly singlet-like so it can easily escape the constraints
from negative results of Higgs searches at the LEP, Tevatron and the LHC. In this case, the next-to-lightest CP-even Higgs boson $h_2$ is the SM-like of mass around 125 GeV.

 The physical spectrum, assuming the CP conserving NMSSM, contains 
 three neutral scalar Higgses $h_{1, 2, 3}$ ($m_{h_1} < m_{h_2} < m_{h_3}$), two pseudoscalars $a_{1, 2}$ ($m_{a_1} < m_{a_2} $) and
a pair of charged Higgses $h^{\pm}$ in addition to five neutralinos $\tilde{\chi}^0_{1, 2, 3, 4, 5}$
($m_{\tilde{\chi}^0_1} < m_{\tilde{\chi}^0_2} < m_{\tilde{\chi}^0_3}< m_{\tilde{\chi}^0_4}< m_{\tilde{\chi}^0_5}$), 
which are mixtures from spin-$\frac{1}{2}$ partners of gauge bosons: the bino ($\tilde{B}^0$), the neutral wino ($\tilde{W}^0$), the
neutral higgsinos ($\tilde{H}^0_d$ and $\tilde{H}^0_u$) and the singlino ($\tilde{S}^0$).\footnote{The charged higgsinos  ($\tilde{H}^+_u$ and $\tilde{H}^-_d$)
 and the charged wino ($\tilde{W}^{\pm}$) also mix with each other, forming the two physical charginos states ($\tilde{\chi}^{\pm}_1$ and $\tilde{\chi}^{\pm}_2$ ).}
 
In the basis ({$\tilde{B}^0$, $\tilde{W}^0$, $\tilde{H}^0_d$, $\tilde{H}^0_u$, $\tilde{S}^0$}), the neutralino-mass matrix is given by

\begin{equation}
 m_{\tilde{\chi}^0_1}
 =
  \begin{bmatrix}
   M\textsubscript{1} &
   0 &
   \frac{g\textsubscript{1}\upsilon\textsubscript{u}}{\sqrt{2}} &
   -\frac{g\textsubscript{1}\upsilon\textsubscript{d}}{\sqrt{2}} &
   0 \\
   0 &
   M\textsubscript{2} &
   -\frac{g\textsubscript{2}\upsilon\textsubscript{u}}{\sqrt{2}} &
   -\frac{g\textsubscript{2}\upsilon\textsubscript{d}}{\sqrt{2}} &
   0 \\
   \frac{g\textsubscript{1}\upsilon\textsubscript{d}}{\sqrt{2}} &
   -\frac{g\textsubscript{2}\upsilon\textsubscript{d}}{\sqrt{2}} &
   0 &
   - \mu_{\rm eff} &
   - \lambda\upsilon\textsubscript{d} \\
   - \frac{g\textsubscript{1}\upsilon\textsubscript{d}}{\sqrt{2}} &
   \frac{g\textsubscript{2}\upsilon\textsubscript{d}}{\sqrt{2}} &
   - \mu &
   0 &
   - \lambda\upsilon\textsubscript{u} \\
   0 &
   0 &
   - \lambda\upsilon\textsubscript{d} &
   - \lambda\upsilon\textsubscript{u} &
   2\kappa\upsilon\textsubscript{s} 
   \end{bmatrix},
\end{equation}
where $M_1$ and $M_2$ are the bino and wino mass parameters, and $g_1$ and $g_2$ are $U(1)_Y$ and $SU(2)_L$ gauge couplings of the SM, respectively.

The neutralino mass eigenstates, as mentioned above, are linear combinations of the bino, the wino, the higgsino and the singlino states,
and can be expressed as:
\begin{equation}
 \tilde{\chi}^0_i = N_{i1}\tilde{B}^0 + N_{i2}\tilde{W}^0 + N_{i3}\tilde{H}^0_d + N_{i4}\tilde{H}^0_u + N_{i5}\tilde{S}^0,
\end{equation}
where the $N_{ij}$'s are obtained by diagonalizing the mass matrix in Eq. (17) with a unitary matrix N,
yielding the five mass eigenstates, ordered by mass. It is clear from Eq. (17) that the NMSSM neutralino sector at tree level is described
by the following input parameters: 
\begin{center}
 $M_1$, $M_2$, $\lambda$, $\kappa$, tan$\beta$ and $\mu_{\rm eff}$ .
\end{center}

We assume that the lightest neutralino, $\tilde{\chi}^0_1$ , is the lightest supersymmetric particle (LSP) and acts as 
a dark matter candidate, which is the focus of this paper.
The bino, wino, higgsino and singlino components of the LSP are defined as $ N^2_{11}$, $ N^2_{12}$, $ N^2_{13} + N^2_{14}$ and  $ N^2_{15}$, respectively. 
The sum of all these components is given by
\begin{equation}
\sum N^2_{1j} = 1.
\end{equation}
In the scenario that the LSP is a bino-like
\begin{equation}
 N_{11}\approx 1,  \hspace{5 mm} N_{15}\approx 0, \hspace{5 mm} N_{13}\approx \frac{m_Z sin\theta_W}{\mu_{\rm eff}}sin\beta, \hspace{5 mm} N_{14}\approx -\frac{m_Z sin\theta_W}{\mu_{\rm eff}}cos\beta.
\end{equation}
In the scenario that the LSP is a singlino-like
\begin{equation}
 N_{11}\approx 0,  \hspace{5 mm} N_{15}\approx 1, \hspace{5 mm} N_{13}\approx -\frac{\lambda\upsilon}{\mu_{\rm eff}}cos\beta, \hspace{5 mm} N_{14}\approx -\frac{\lambda\upsilon}{\mu_{\rm eff}}sin\beta.  
\end{equation}

In the context of the NMSSM, the lightest neutralino $\tilde{\chi}^0_1$ can be bino-like, singlino-like and higgsino-like.
In some cases, the $\tilde{\chi}^0_1$, the second-lightest neutralino $\tilde{\chi}^0_2$, and the third neutralino $\tilde{\chi}^0_3$ are almost degenerate
with nearly the same mass. The bino-like neutralino requires that $\mu_{\rm eff} >> M_1$. The higgsino-like neutralino can be obtained when 
the ratio of the absolute value of $\kappa$ and $\lambda$ is quite large in contrast to the singlino-like neutralino 
which requires that $\kappa/\lambda$ is quite small, as can be seen from the neutralino mixing matrix coefficient
$2\kappa\upsilon\textsubscript{s} = 2(\kappa/\lambda)\mu_{\rm eff}$. The higgsino content of the lightest neutralino
is important for the elastic scattering of the lightest neutralino with a nucleus.

The $\tilde{\chi}^0_1$ as a dark matter candidate might be directly detected in deep underground experiments by measuring its recoil
 off a nucleus. This direct scattering  can be divided into two types: spin-independent (SI)
 and spin-dependent (SD). In the latter one, the $\tilde{\chi}^0_1$ couples to the spin of the nucleus.
 
 The SI cross section of the $\tilde{\chi}^0_1$-nucleon scattering is dominated by
 a t-channel exchange of the scalar Higgs bosons. This cross section is proportional to the higgsino content of the $\tilde{\chi}^0_1$ ($\sigma^{SI} \propto N^2_{13} + N^2_{14}$)
 and also to the mass squared of the nucleus ($m^2_{\tilde{\chi}^0_1}$), which yields a substantial enhancement for heavy
nuclei \cite{Drees:1993bu}. On the other hand, the dominant diagram for the elastic SD scattering is the $Z^0$ boson exchange. The corresponding SD cross section
is proportional to the difference of the higgsino components of the $\tilde{\chi}^0_1$ ($\sigma^{SD} \propto |N^2_{13} - N^2_{14}|$).

For the relic density calculation, we demand the calculated
relic density corresponding to the measured data $\Omega h^2$ = 0.1187 from the Planck satellite experiment in addition to $10$\% theoretical uncertainty.
So, we choose the relic density requirement of the lightest neutralino as
\begin{equation}
 0.10683 < \Omega h^2 < 0.13057.
\end{equation}

The MSSM limit of NMSSM is obtained by taking both $\lambda \to 0$ and $\kappa \to 0$ while keeping  $\mu_{\rm eff} \neq 0$, which requires
a large $\upsilon\textsubscript{s}$. This leads to decoupling the singlet superfield from interacting with the Higgs doublet superfields.
In such a case, the mixing of the singlino with other neutralinos vanishes,
which makes it impossible to distinguish the NMSSM neutralino sector from the one of the MSSM. This mixing
is controlled by the $\lambda$ parameter. In the limit $\lambda \to 0$, the singlino is almost a pure state with mass
\begin{equation}
 m_{\tilde{S}^0} = 2\kappa\upsilon\textsubscript{s}. 
\end{equation}
When the $\tilde{\chi}^0_1$ has a tiny singlino component, coupling to the scalar Higgses requires the $ \tilde{\chi}^0_1$ of a mixture of bino and higgsino,
as in the MSSM.

A characteristic feature of the NMSSM is the possibility of having a singlino-like dark matter $\tilde{\chi}^0_S$ , being LSP,
but due to the small couplings to SM particles, it is
not-trivial to obtain correct dark matter density for a nearly $ \tilde{\chi}^0_S$. For almost singlino DM in the scenario of very small values of
both $\lambda$ and $\kappa$, no efficient annihilation mechanisms are available.
However, the correct relic density of an NMSSM LSP with WMAP measurements can be obtained by annihilation of two of
the next-to-lightest supersymmetric particles NLSP-NLSP annihilation
(assisted coannihilation) \cite{Djouadi:2008uj}. Further, in this scenario
all supersymmetric particle decay chains proceed as in the MSSM into NLSP, which will decay into $\tilde{\chi}^0_S$ + SM particles if
kinematically allowed. Due to a very small NLSP-$\tilde{\chi}^0_S$ coupling, the NLSP can have a very long life time leading to displaced
vertices \cite{Ellwanger:1997jj}. Then, this decay allow to distinguish the
NMSSM from the MSSM in this case. Assuming that the NLSP is the second-lightest neutralino $\tilde{\chi}^0_2$,
it can decay in many different ways if kinematically allowed, such as $\tilde{\chi}^0_2\to \tilde{\chi}^0_S Z$, $\tilde{\chi}^0_2\to \tilde{\chi}^0_S \gamma$,
$\tilde{\chi}^0_2\to \tilde{\chi}^0_S l^+ l^-$, $\tilde{\chi}^0_2\to \tilde{\chi}^0_S \nu \bar{\nu}$, $\tilde{\chi}^0_2\to \tilde{\chi}^0_S q \bar q$,
and $\tilde{\chi}^0_2\to \tilde{\chi}^0_S \Phi$, where $\Phi$ is a scalar or pseudoscalar Higgs.

\section{Scanning of parameter space }
\label{sect:scan1}

In this section, we try to identify the NMSSM parameter space in which the lightest neutralino is the LSP. The samples of the parameter space must satisfy relic
density, theoretical and experimental constraints as well as direct and indirect detection
limits. We use the code NMSSMTools (v5.6.1) \cite{NMHDECAY1,NMHDECAY2,NMSSMTools} to calculate the
mass spectrum of the supersymmetric particles from the NMSSM parameters. This code also 
calculates the masses, couplings, and decay widths of all the 
Higgs bosons, taking into account theoretical and experimental constraints. All theoretical and experimental constraints are
considered here apart from the muon anomalous magnetic moment.\footnote{The effect of this constraint on the NMSSM
parameter space was discussed in \cite{Domingo:2008bb}.} Theoretical constraints include no Landau pole is developed in the running $\lambda$, $\kappa$ and Yukawa couplings
below the grand unified scale and the physical vacuum is the global minimum. 
Radiative electroweak symmetry breaking requires the couplings $\kappa^2 < \lambda^2$
at the unification scale.
The following constraint $\sqrt{\lambda^2 + \kappa^2} \leq 0.7$ 
is applied at the electroweak scale to maintain a perturbable theory up to the grand unification scale.
The experimental constraints include limits on upsilon, B and K decays.
Further, we consider constraints from the direct searches for Higgs bosons and supersymmetric particles at LEP, Tevatron and the LHC as implemented in the code.
We also require the SM-like Higgs mass to be  within the range of 122--128 GeV to allow for theoretical uncertainties, with its production rates fitting
the LHC data. 

The dark matter relic density in addition to  both the spin-independent (SI)
 and spin-dependent (SD) DM-nucleon scattering cross sections
are computed with the package MicrOMEGA\textunderscore5.0 \cite{Belanger:2013oya} integrated with the NMSSMTools.
The observed DM relic abundance is tested with the measured data $\Omega h^2$ = 0.1187 at the 
 $10$\% level, see Eq. (22). Limits on both the spin-independent (SI)
 and spin-dependent (SD) DM-nucleon scattering cross sections
from LUX, XENON1T and PICO-60 are taken into account \cite{LUX:2017ree,XENON:2018voc,XENON:2019rxp,PICO:2019vsc}.

In our parameter space, we perform a scan over the following ranges:
\begin{center}
$50 \leq M_1 \leq 2000$,  \phantom{aa} $100 \leq M_2 \leq 3000$, \phantom{aa} $100 \leq M_3 \leq 5000$,
\phantom{aa} $0.0001\leq \lambda \leq  0.1$, \phantom{aa} $-0.1 \leq \kappa \leq  0.1$,
 \phantom{aa} $1.6 \leq \tan\beta \leq  60$, \phantom{aa} \\
$100 \leq \mu_{\rm eff} \leq  1000$ GeV, \phantom{aa} $-5000 \leq A_{\lambda} \leq  5000$ GeV,\phantom{aa} $-5000 \leq A_{\kappa} \leq  5000$ GeV. \\
\end{center}
Here, we consider the case of small values of $\lambda$ and $\kappa$ to make our results more specific to study
the properties of the dark matter candidate in the NMSSM parameter space, and to see if the results are different
from those of the MSSM in the neutralino sector or not, since there
are only two terms different in the superpotential between
the NMSSM and the MSSM, see Eq. (1).
Remaining soft mass parameters for squarks and sleptons as well as
the trilinear soft SUSY coupling parameters, contributing at higher order level, are set to \\
$\bullet\phantom{a}m_Q = m_U = m_D = m_L = m_E = m_{Q_3} = m_{U_3} = m_{D_3} = m_{L_3} = m_{E_3} = 2000$ GeV,\\
$\bullet\phantom{a}A_{U_3} = A_{D_3} = A_{E_3} = 5000$ GeV.\\

To determine the parameter space compatible with all current theoretical and
experimental constraints mentioned above, we scan five million random points in the
specified parameter space with the input parameters specified at the SUSY breaking scale except tan$\beta$ at $m_Z$. The points selected for the scan are with linear-flat distributions over the specified
range of the parameters. The outcome of the scan contains masses of the Higgs bosons and the supersymmetric particles, both spin-independent (SI) and spin-dependent
(SD) cross sections of DM-nucleon scattering and the relic density for all the surviving data points which have passed 
the constraints. The points which violate the above constraints are eliminated by the code.
The surviving data points are then used in our numerical results.

\section{Results and discussions}

In this section, we present the numerical results from our analysis to study characteristics of the dark matter candidate and its properties.
Figs. 1 and 2 show how the lightest neutralino mass  $m_{\tilde{\chi}^0_1}$ varies with respect to 
the relevant input parameters in addition to  the ratio of $\kappa$ and $\lambda$ multiplied by $\mu_{\rm eff}$, $(|\kappa/\lambda|)\mu_{\rm eff}$, which determine the physics of the neutralino sector.
It is evident from the figures that the parameters: $\lambda$, tan$\beta$, $A_{\lambda}$ and $A_{\kappa}$ (and,
to some extent, also large $M_1$ and $M_2$) have quite uniform distributions.
It is also clear from these figures that $\tilde{\chi}^0_1$ masses with about 60 GeV and 1000 GeV are preferred,
since the mass depends strongly on the $(|\kappa/\lambda|)\mu_{\rm eff}$ as shown in the bottom panel of Fig. 2.
As for the remaining two parameters $\kappa$ and $\mu_{\rm eff}$, it is noticeable that they have different distribution
patterns than the others. For $\mu_{\rm eff}$, a linear relationship with $m_{\tilde{\chi}^0_1}$ appears in which
the surviving points are mostly concentrated, whereas $\kappa$ behaves a little
differently where most of the sample points clustering in the range $|\kappa| \lesssim 0.4 $ rather than being distributed randomly. This has led to the conclusion that 
the parameters that have the most significant effects on the mass of the lightest neutralino are $\kappa$ and $\mu_{\rm eff}$.

Fig. 3 shows the mass components of the lightest neutralino as a function of its mass. The upper left panel of the figure shows
the bino component, $N^2_{11}$, of $\tilde{\chi}^0_1$. It is clear that its contribution to the $m_{\tilde{\chi}^0_1}$ is negligible in most of our
parameter space with the possibility that the $N^2_{11}$ is dominant, reaching unity in small region of the parameter space with $m_{\tilde{\chi}^0_1} \lesssim 725$ GeV
in which the $\tilde{\chi}^0_1$ is  bino-like. The latter region occurs for $-5 \lesssim \lambda/\kappa \lesssim 5$ and  $\mu_{\rm eff} \gtrsim 400$ GeV, see the middle and right 
upper panels of Fig. 4. The upper right panel of Fig. 3 shows that the wino composition in $m_{\tilde{\chi}^0_1}$ is always small in our parameter space.
 The lower panels of Fig. 3 show the higgsino (left) and singlino (right) fractions in $m_{\tilde{\chi}^0_1}$.
 One can generally observe from these two panels that the higgsino component increases by increasing $m_{\tilde{\chi}^0_1}$ whereas
 the singlino component decreases by increasing $m_{\tilde{\chi}^0_1}$. It is also clear that for most of the surviving points of our parameter space,
 the LSP dark matter is singlino-dominated, in contrast to the MSSM where the lightest neutralino is mostly bino-dominated.
 The dominance of the singlino contribution $ N^2_{15}\approx 1$ occurs
in the region of $-0.4 \lesssim \kappa  \lesssim 0.4 $ and either $\lambda/\kappa \lesssim -2.5$ or $\lambda/\kappa \gtrsim 2.5$, see the left and middle lower panels of
Fig. 4, while the dominance of the higgsino contribution $N^2_{13} + N^2_{14}\approx 1$ occur in the region of  $-2.5 \lesssim \lambda/\kappa \lesssim 2.5$
and $\mu_{\rm eff} \gtrsim 800$ GeV as shown in the middle panels of the figure.
Briefly, we have noticed that for the largest values of $m_{\tilde{\chi}^0_1}$, the $\tilde{\chi}^0_1$ is higgsino-like
whereas for the smallest values of $m_{\tilde{\chi}^0_1}$, the $\tilde{\chi}^0_1$ is either singlino-like or bino-like. 

Following are the features of the surviving samples classified
according to the dominant component of dark matter. 

1- Bino-dominant dark matter scenario \\
For this scenario, the bino dominated neutralino LSP requires moderate or large values of $\mu_{\rm eff}$, and its mass $m_{\tilde{\chi}^0_1}$ is approximately given
by the bino mass parameter $M_1$. In contrast to the MSSM, light bino-dominated dark matter can have 
a relic density consistent with WMAP results due to a light CP-odd Higgs state
which can be exchanged in the s-channel.

2- Singlino-dominant dark matter scenario \\
 The singlino dominated neutralino LSP depends on the ratio of the absolute value of $\kappa$
and $\lambda$, since $m_{\tilde{S}^0} = 2\kappa\upsilon\textsubscript{s}$. Such a scenario is possible for values of $\lambda/\kappa$ 
below $\sim$ -2.5 or above $\sim$ 2.5. If the neutralino LSP is predominantly a singlino, it may have suppressed coupling to any SM particle
and thus small scattering cross section. In this case, a light neutralino with a large singlino component may evade detection in future experiments.
 Further, the singlino-dominated LSP can cover a large range of relic densities, since
many co-annihilation channels, such as the $\tilde{\chi}^0_1$ coannihilated with the higgsino-dominated neutralinos and
charginos, can contribute. So, it is easy to fulfill the correct dark matter density in this scenario.

3- Higgsino-dominant dark matter scenario \\
The higgsino dominated neutralino LSP requires large values of $\mu_{\rm eff}$ (above ∼ 800 GeV). In this scenario
the dark matter is mainly annihilated into $W^+W^-$ and $ZZ$. In fact, the higgsino content 
is crucial for the elastic scattering cross section, since
this maily proceeds through the exchange of a Higgs boson. 

 The upper left (right) panel of Fig. 5 shows the SI $\tilde{\chi}^0_1$ -proton
 (-neutron) cross-section $\sigma^{SI}_p (\sigma^{SI}_n)$ 
 versus the $m_{\chi_1^0}$  while  the lower left (right) panel of the figure shows
 the SD $\tilde{\chi}^0_1$ -proton (-neutron) cross-section $\sigma^{SD}_p (\sigma^{SD}_n)$
  versus the $m_{\chi_1^0}$. 
 It is quite evident that the cross sections of the proton and neutron have similar behavior, but they have
 different values as we have noticed that $\sigma^{SI}_p \approx  1.02$ $\sigma^{SI}_n$ and $\sigma^{SD}_p \approx  0.76$ $\sigma^{SD}_n$. 
 Almost all the surviving points have the $\sigma^{SI}$ and $\sigma^{SD}$ of the order $10^{-9}$ and $10^{-5}$ pb, respectively, 
or smaller, below the current limits from various direct detection experiments of WIMP dark matter \cite{LUX:2017ree,XENON:2018voc,XENON:2019rxp,PICO:2019vsc}.
 Fig. 6 shows the higgsino component $N^2_{13} + N^2_{14}$ of the $\tilde{\chi}^0_1$ versus both $\sigma^{SI}_p$ and $\sigma^{SI}_n$ (upper-panels) and 
 the difference of the higgsino components $|N^2_{13} - N^2_{14}|$ of the $\tilde{\chi}^0_1$ versus both $\sigma^{SD}_p$ and $\sigma^{SD}_n$ (bottom-panels).
 It is clear that the larger the higgsino component of the $\tilde{\chi}^0_1$, the larger $\sigma^{SI}$ and also
 the larger the difference of the higgsino components of the $\tilde{\chi}^0_1$, the larger $\sigma^{SD}$.

 Figs. 7 and 8 present the correlations between the second-lightest neutralino mass, $m_{\tilde{\chi}^0_2}$, third neutralino mass, $m_{\tilde{\chi}^0_3}$,
lightest chargino mass, $m_{\tilde{\chi}^{\pm}_1}$ and the lightest neutralino mass, $m_{\tilde{\chi}^0_1}$ (Fig. 7) and
the correlations between the lightest and second-lightest neutral scalar Higgs boson masses, $m_{h_1}$ and $m_{h_2}$, the lightest pseudo-scalar
neutral Higgs mass, $m_{a_1}$ and $m_{\tilde{\chi}^0_1}$ (Fig. 8). It is obvious that the $m_{\tilde{\chi}^0_1}$, $m_{\tilde{\chi}^0_2}$, $m_{\tilde{\chi}^0_3}$
and  $m_{\tilde{\chi}^{\pm}_1}$ are 
mass-degenerate in most of the allowed parameter space except for small values of $m_{\tilde{\chi}^0_1}$,
$\lesssim 100$ GeV and large values of $m_{\tilde{\chi}^0_3}$, $\gtrsim$ 1000 GeV. 
It is also very clear from the upper left panel of Fig. 8 that the $h_1$ is the SM-like
Higgs boson, $ 122\lesssim m_{h_1} \lesssim 128$ GeV, in our parameter space, which has small values of both $\lambda$ and $\kappa$ and
consistent with the relic abundance considerations in addition to the direct and indirect searches for dark matter. Further, the $a_1$ is
a singlet-like state with mass between $77 - 87$ GeV.
The discovery of such a very light pseudo-scalar Higgs state with $m_{a_1} \lesssim m_Z$ at the LHC or other colliders
would unmistakably indicate the existence of the non-minimal
nature of the SUSY Higgs sector, see e.g. Ref. \cite{Almarashi:2021pgu} (and references therein).
 
 Ultimately, computing the relic density of the lightest neutralino as a function of the $m_{\tilde{\chi}^0_1}$ in Fig. 9 give us
 a clear indication of our parameter space with the correct relic cold dark matter,
 which is mostly possible for a singlino-dominated neutralino as the LSP.
The sample points that are not consistent with the relic density considerations have been excluded. One can see that
these right amount of the dark matter can be fulfilled for different values of $m_{\tilde{\chi}^0_1}$, starting
from about 60 GeV up to about 1 TeV and for any neutralino composition, whether it is singlino-like, higgsino-like, bino-like or
a mixture of them.

Finally, we briefly discuss the latest constraints from the LHC on the chargino-neutralino (electroweakinos) sectors.
Dedicated LHC searches targeting $\tilde{\chi}^0_2$ and $\tilde{\chi}^{\pm}_1$ production have put
stringent lower bounds on their masses \cite{ATLAS:2019lff,ATLAS:2020pgy,ATLAS:2019wgx,CMS:2021cox}. These states can decay to the lightest neutralino
in addition to $h_{SM}/W/Z$. A Search by ATLAS collaboration using 
data collected with approximately 139 fb$^{-1}$ of integrated luminosity has excluded gluino with masses up to 2.2 TeV for a bino-like $\tilde{\chi}^0_1$
with $m_{\tilde{\chi}^0_1} \lesssim$ 200 GeV \cite{ATLAS:2021twp}.
However, such constraints could be easily invalidated in the NMSSM in the presence of the singlino,
and thus weakening the lower bounds on the mass of such a neutralino.

\begin{figure}
 \centering\begin{tabular}{cc}
 \includegraphics[scale=0.5]{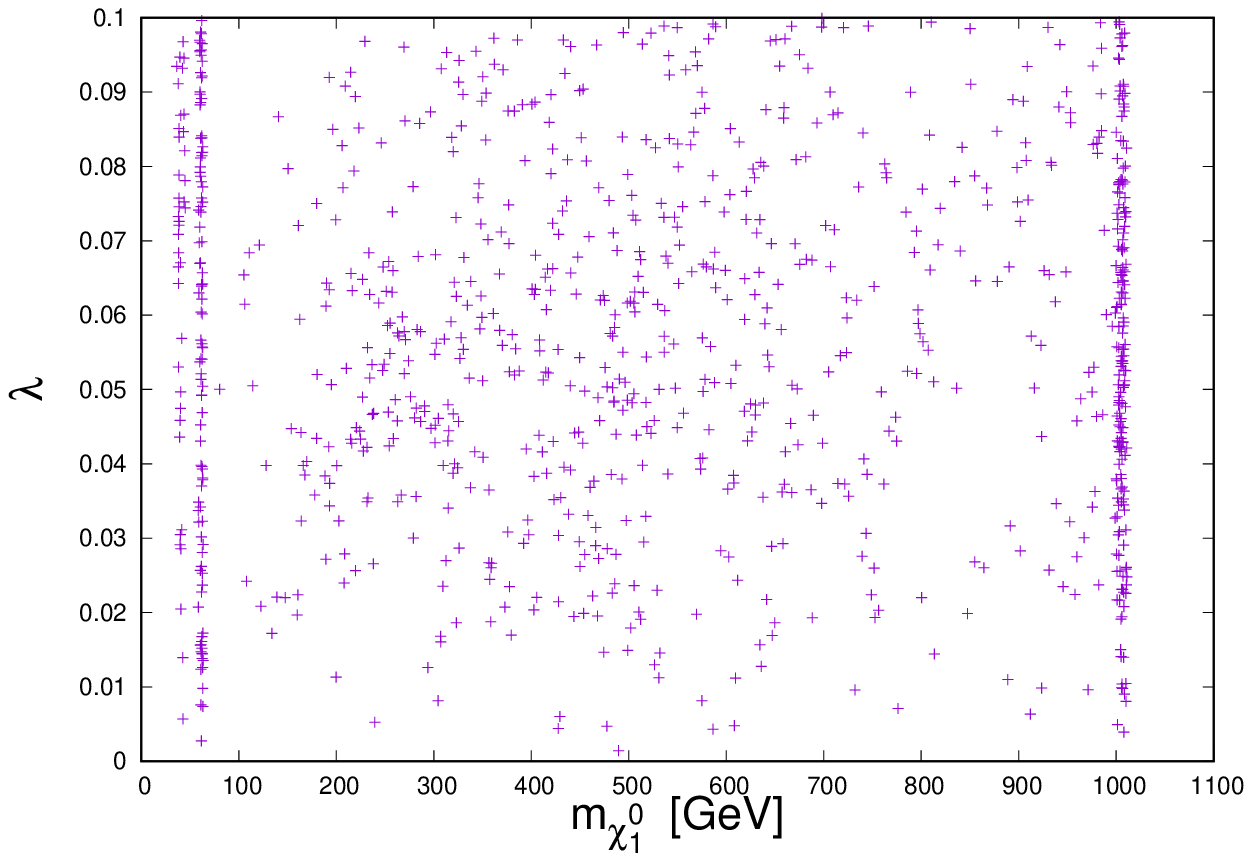} &\includegraphics[scale=0.5]{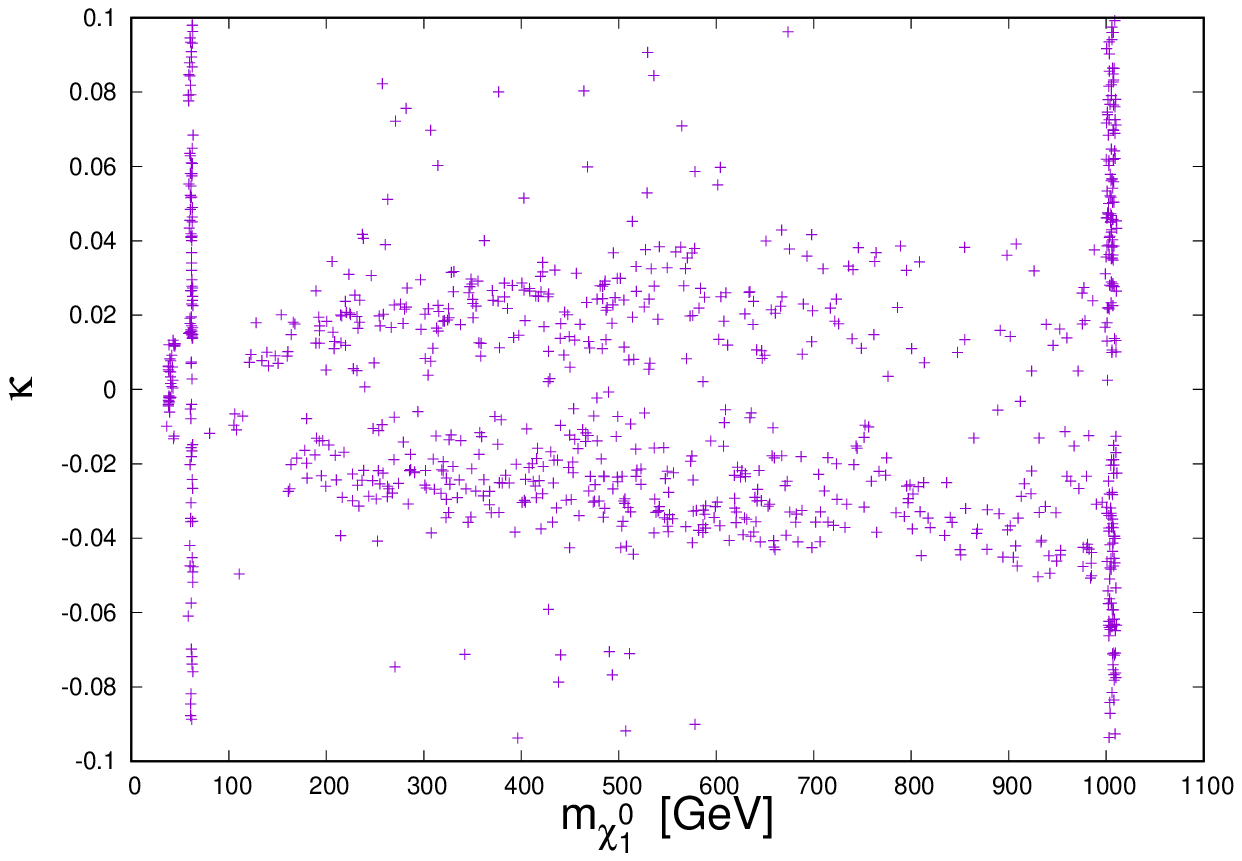}\\
 \includegraphics[scale=0.5]{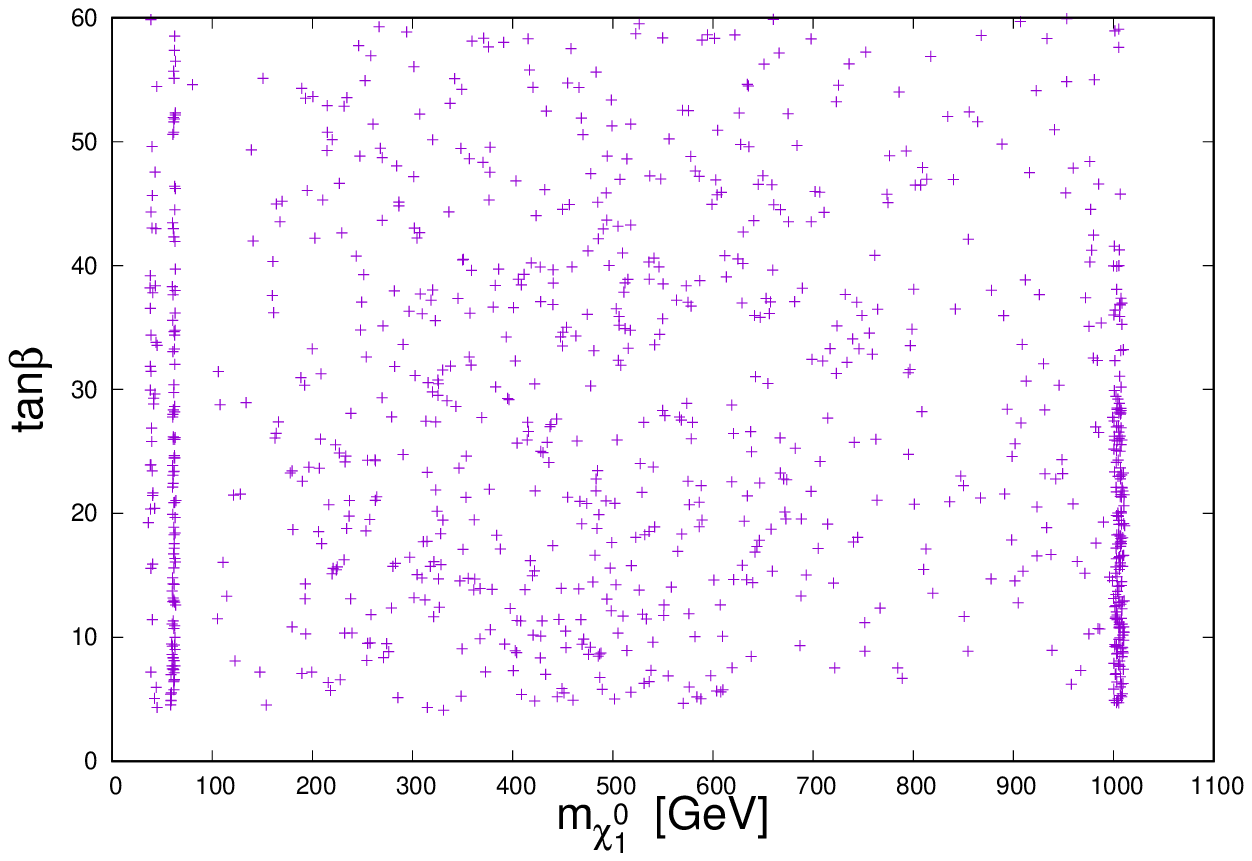} &\includegraphics[scale=0.5]{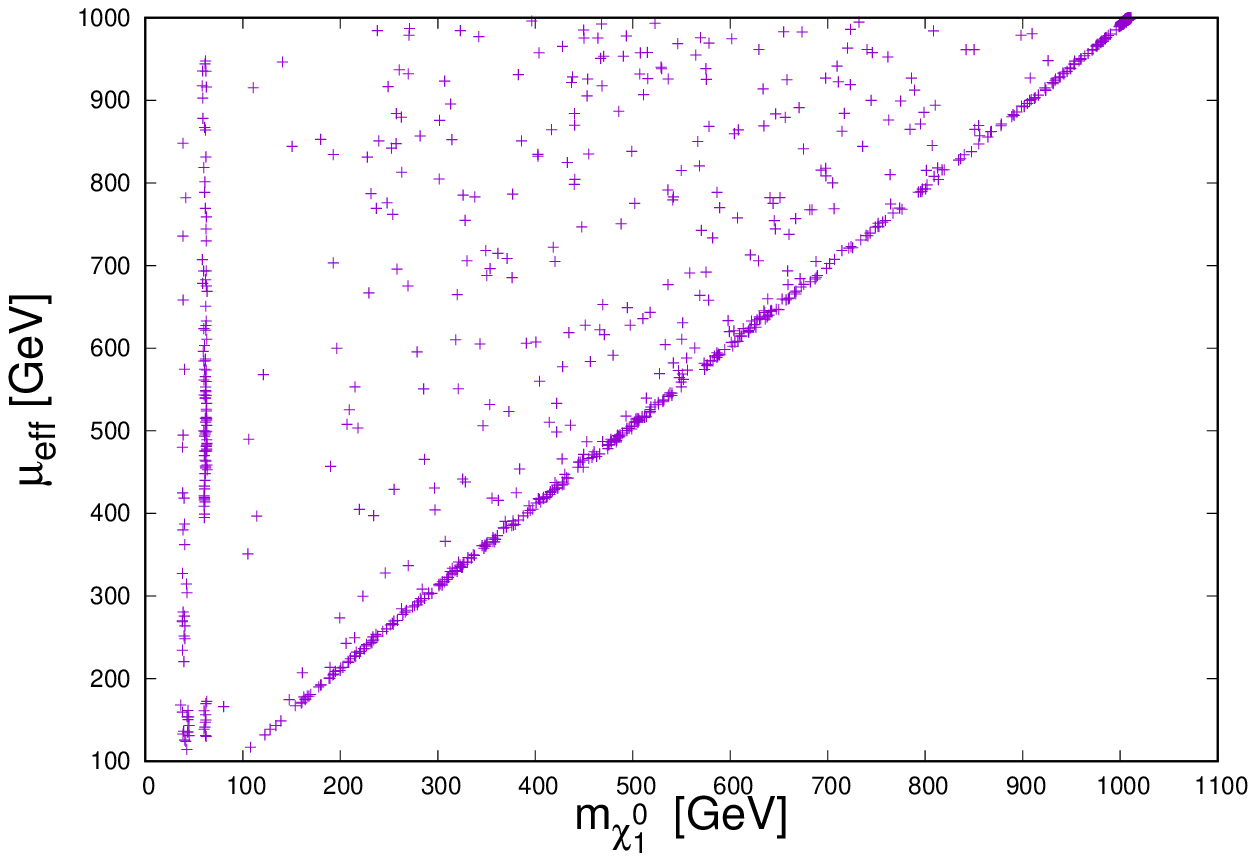}\\
 \includegraphics[scale=0.5]{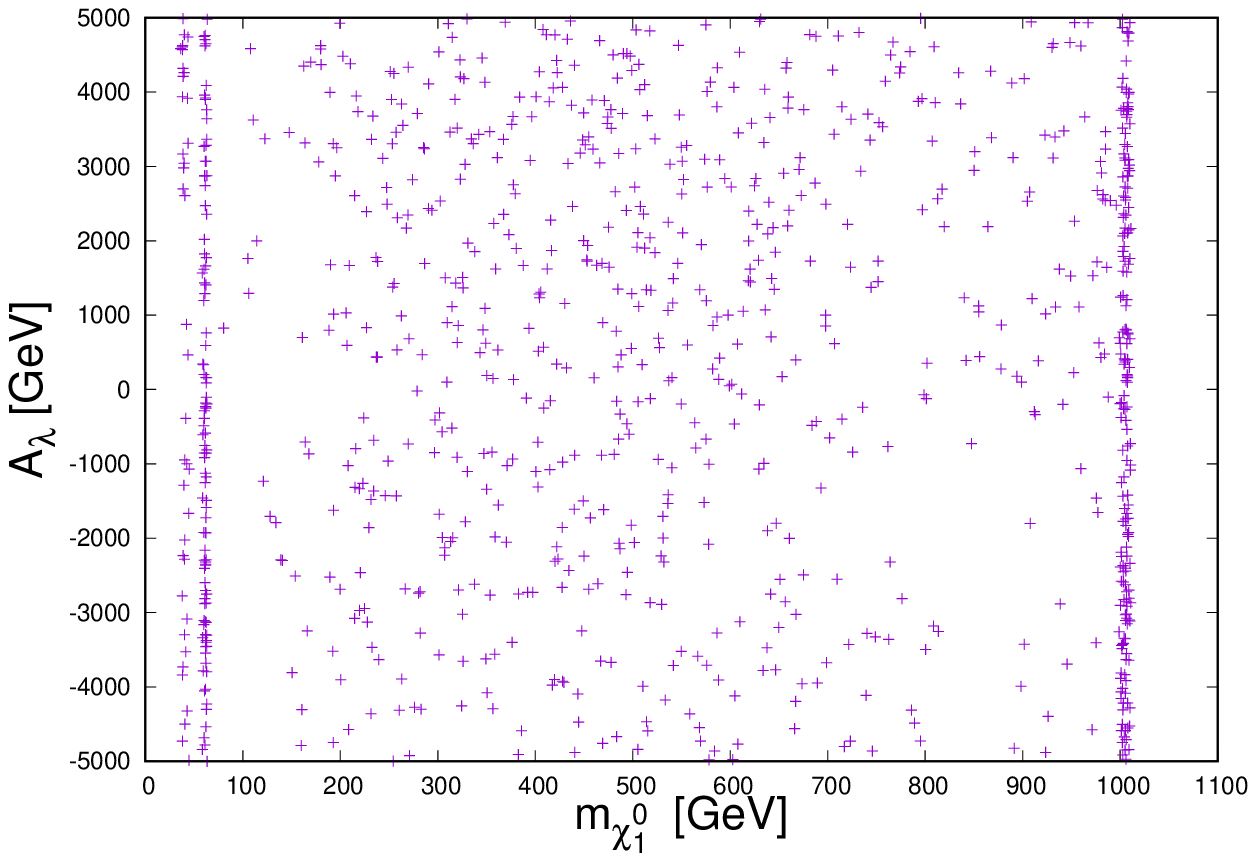} & \includegraphics[scale=0.5]{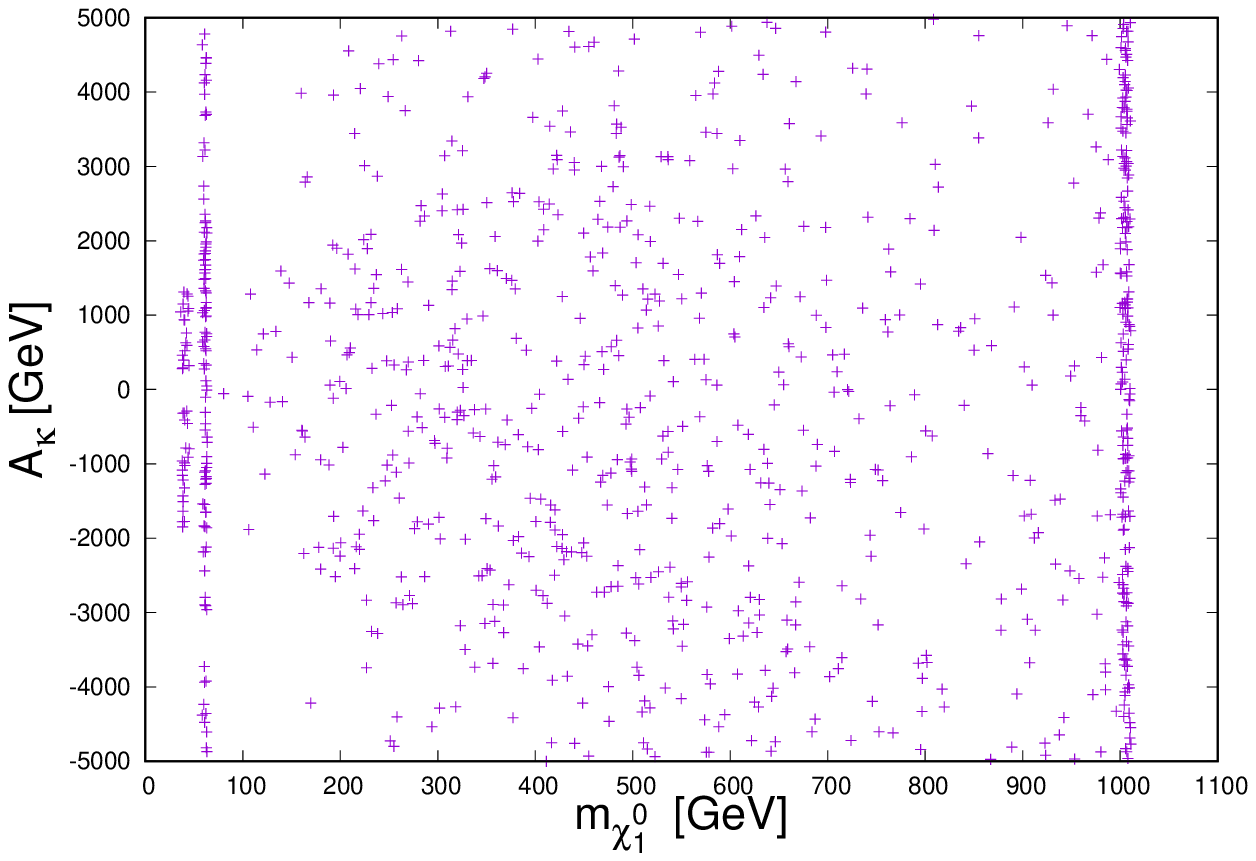}
 \end{tabular}

\caption{The lightest neutralino mass $m_{\tilde{\chi}^0_1}$ as a function of $\lambda$ and $\kappa$ (top-panels), $\tan\beta$ and $\mu_{\rm eff}$ (middle-panels)
and,  $A_\lambda$ and $A_\kappa$ (bottom-panels).}
  \label{fig1}
\end{figure}

\begin{figure}
 \centering\begin{tabular}{cc}
 \includegraphics[scale=0.5]{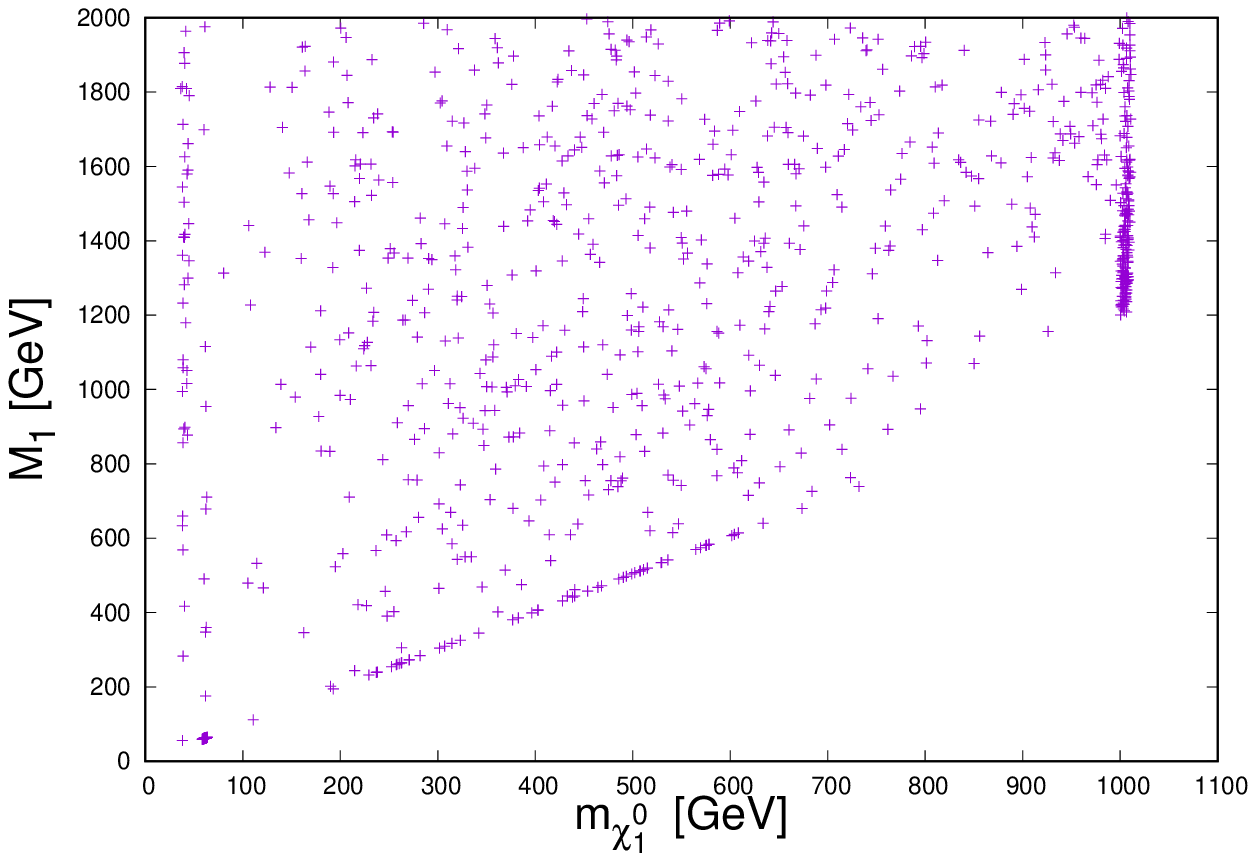} & \includegraphics[scale=0.5]{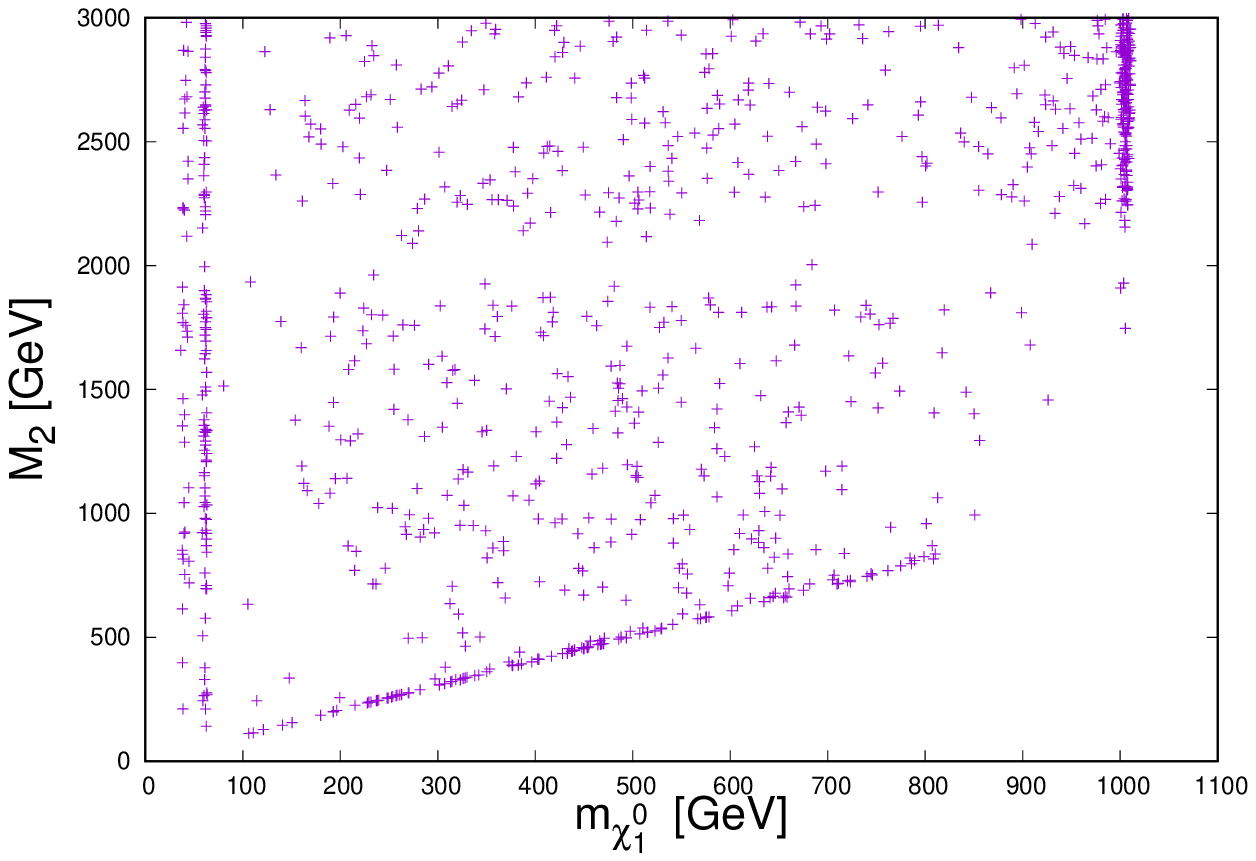} \\
  \end{tabular}
 \centering
  \includegraphics[scale=0.5]{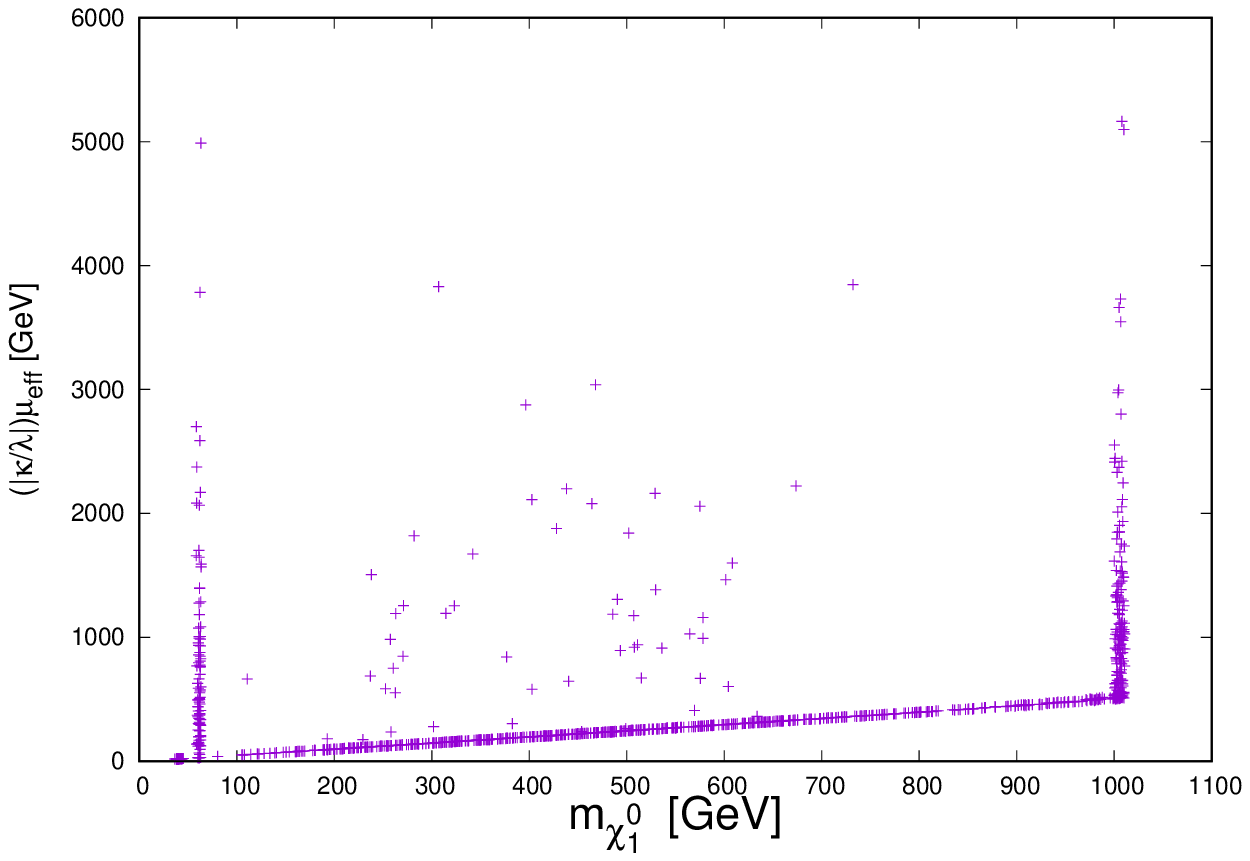}

\caption{The lightest neutralino mass $m_{\tilde{\chi}^0_1}$
versus the bino and wino mass parameters $M_1$ and $M_2$ (upper panels), and versus $(|\kappa$/$\lambda|)\mu_{\rm eff}$.}
\label{fig2}
\end{figure}

\begin{figure}
 \centering\begin{tabular}{cc}
  \includegraphics[scale=0.5]{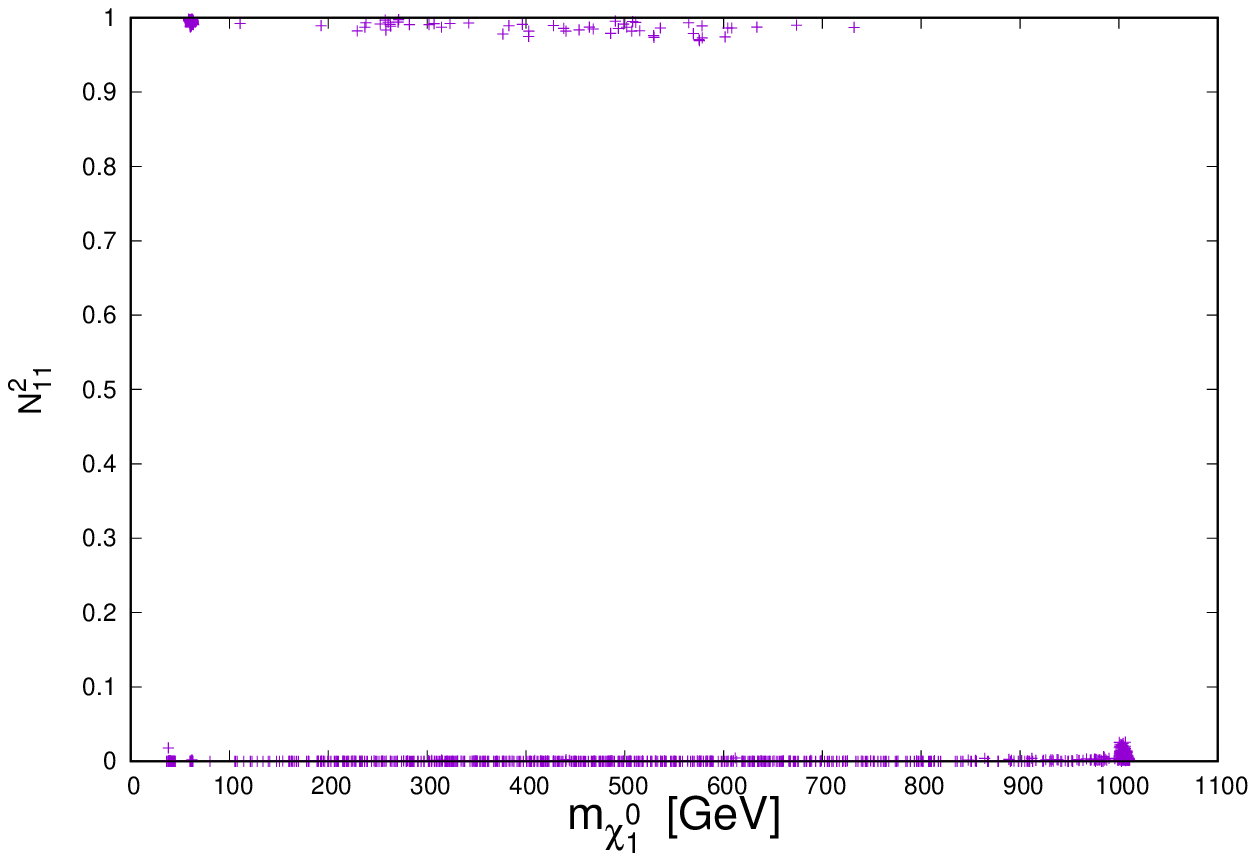} &\includegraphics[scale=0.5]{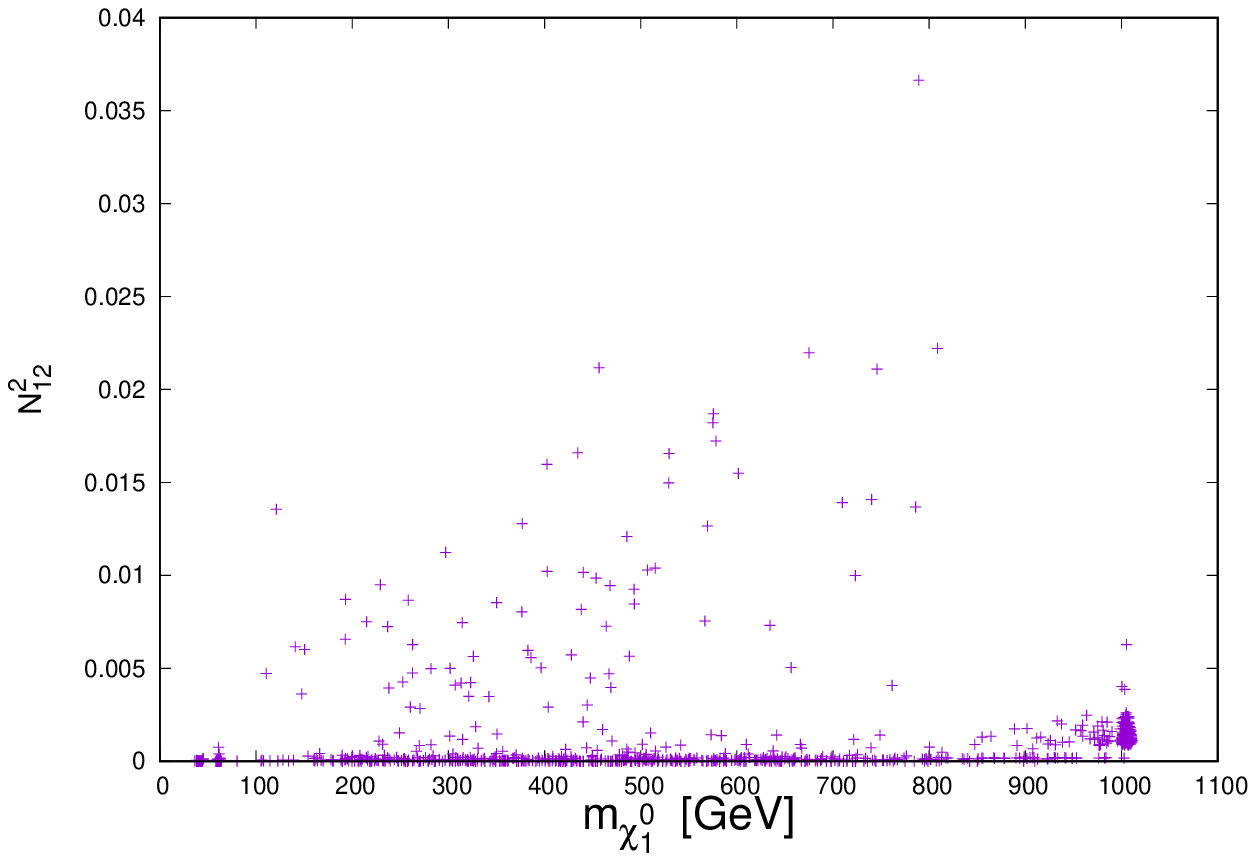}\\
 \includegraphics[scale=0.5]{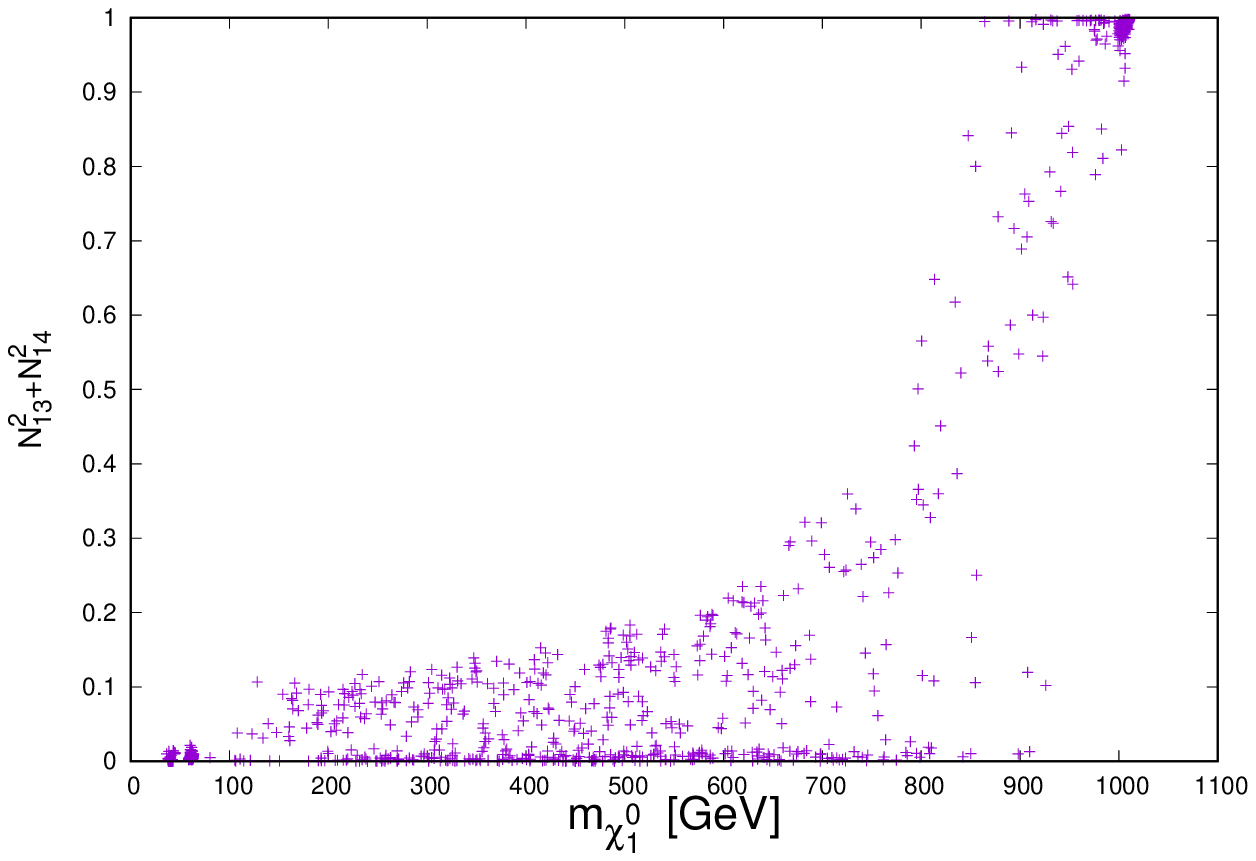} &\includegraphics[scale=0.5]{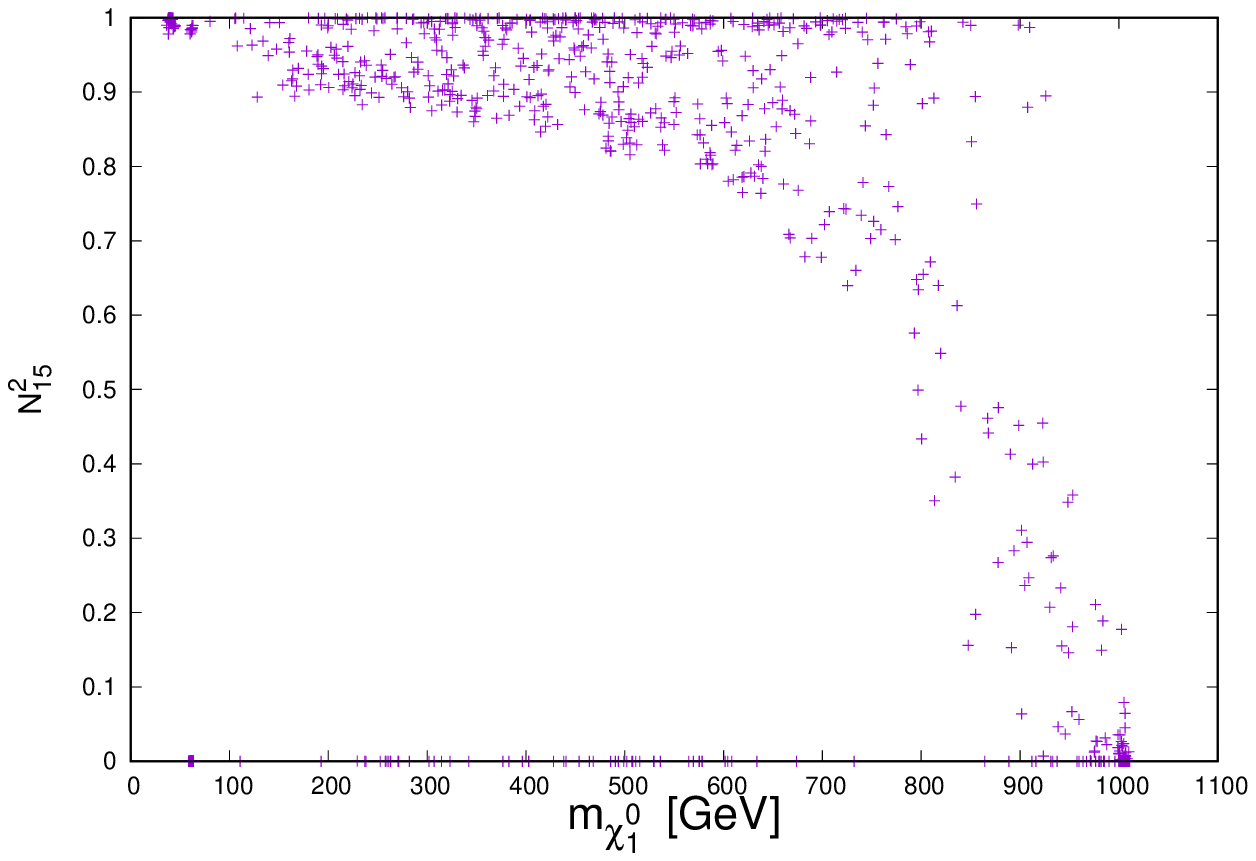} 

 \end{tabular}

\caption{The mass components of the lightest neutralino, $\tilde{\chi}^0_1$. }
\label{fig3}
\end{figure}

\begin{figure}
 \centering\begin{tabular}{ccc}
 \includegraphics[scale=0.33]{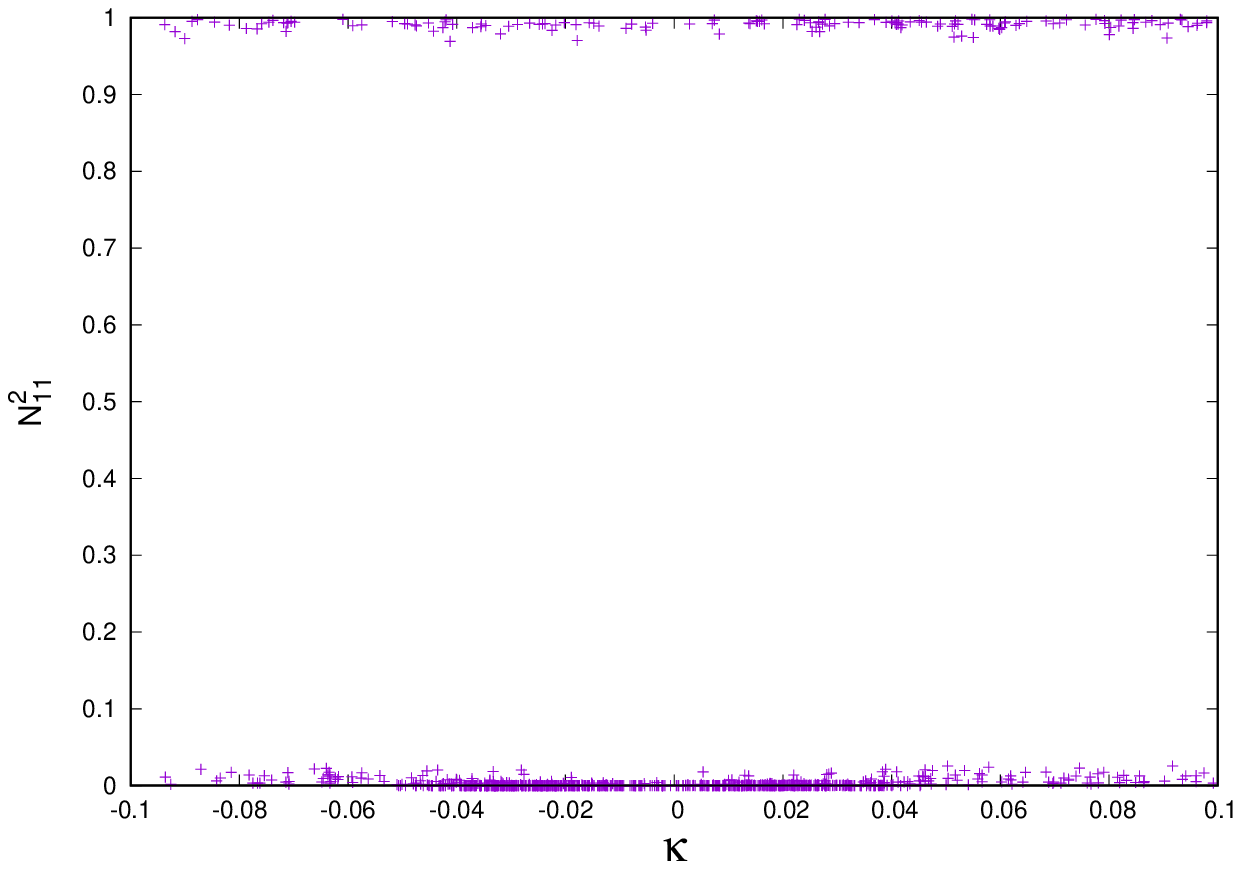} &\includegraphics[scale=0.33]{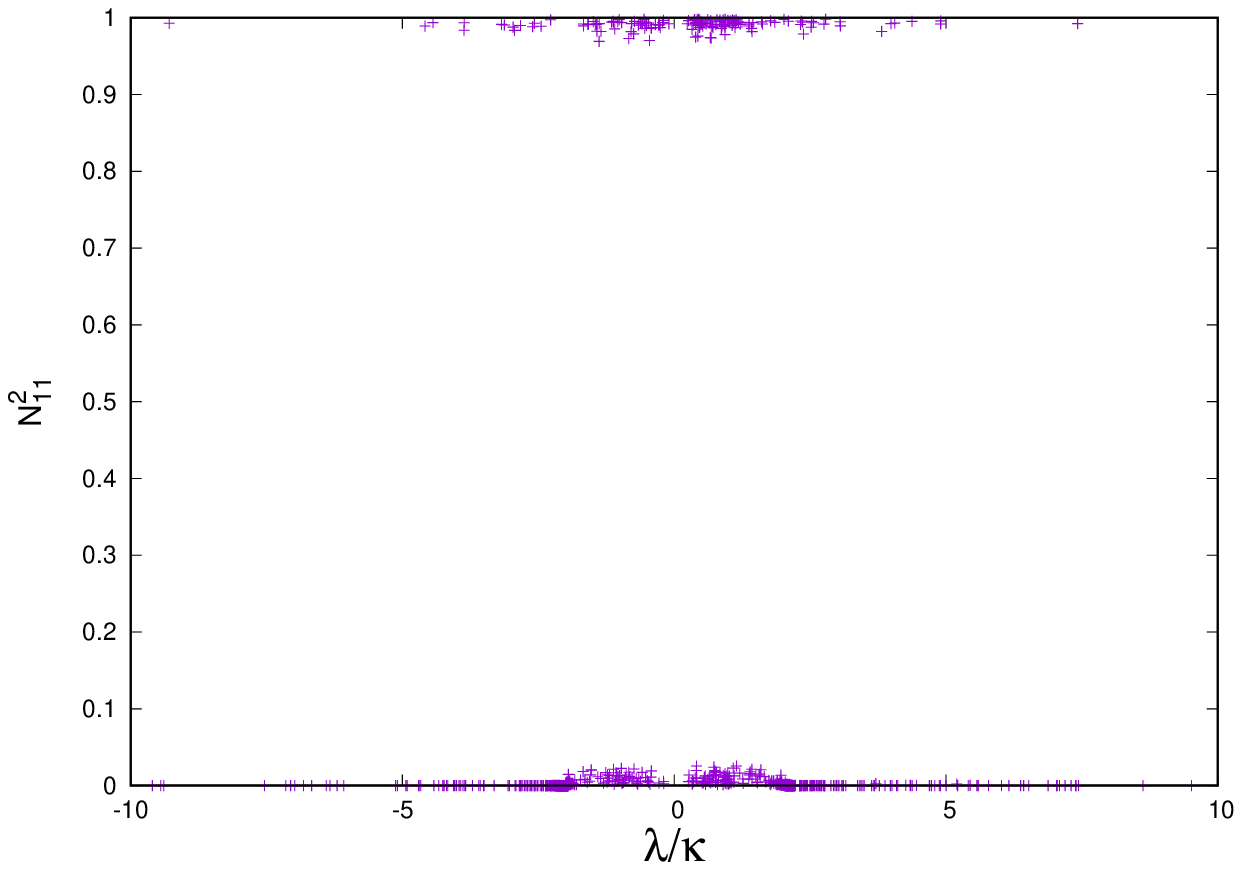}&\includegraphics[scale=0.33]{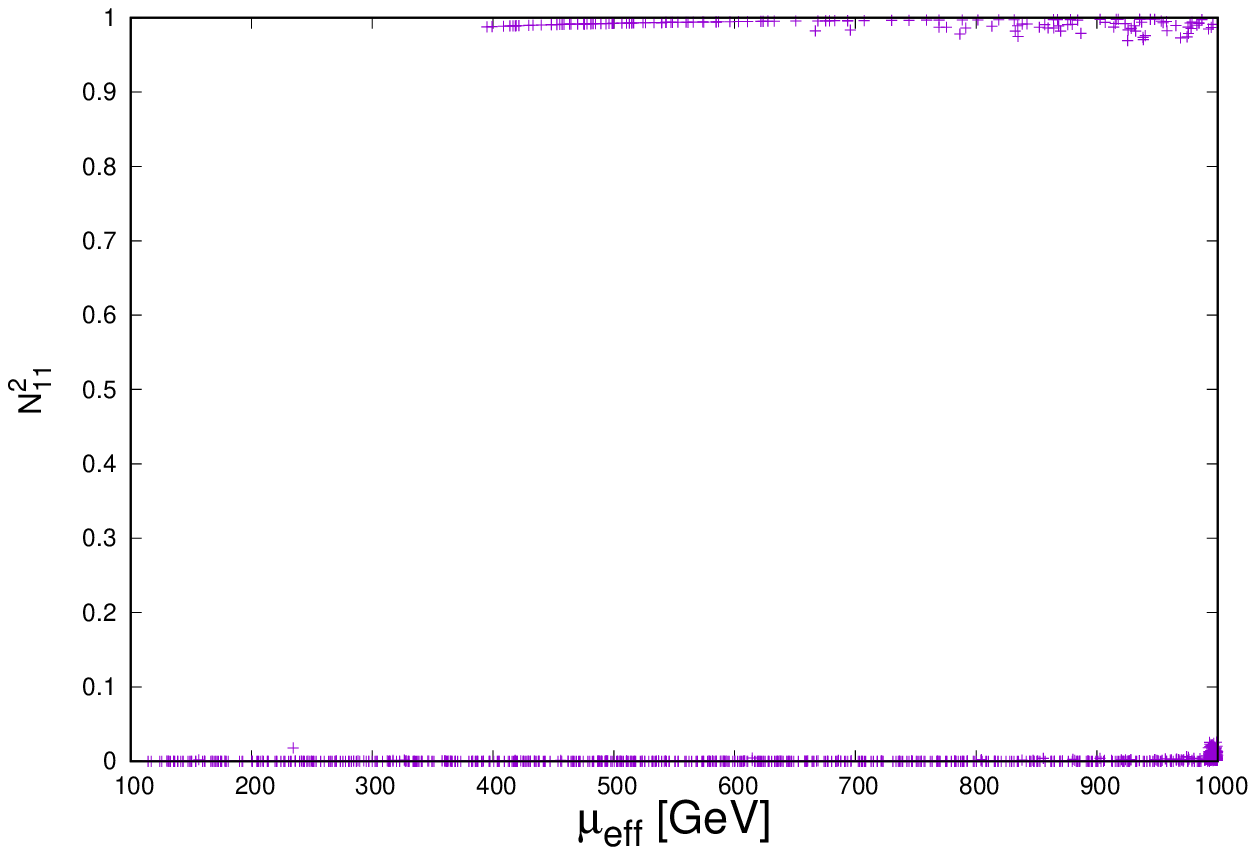}\\
 \includegraphics[scale=0.33]{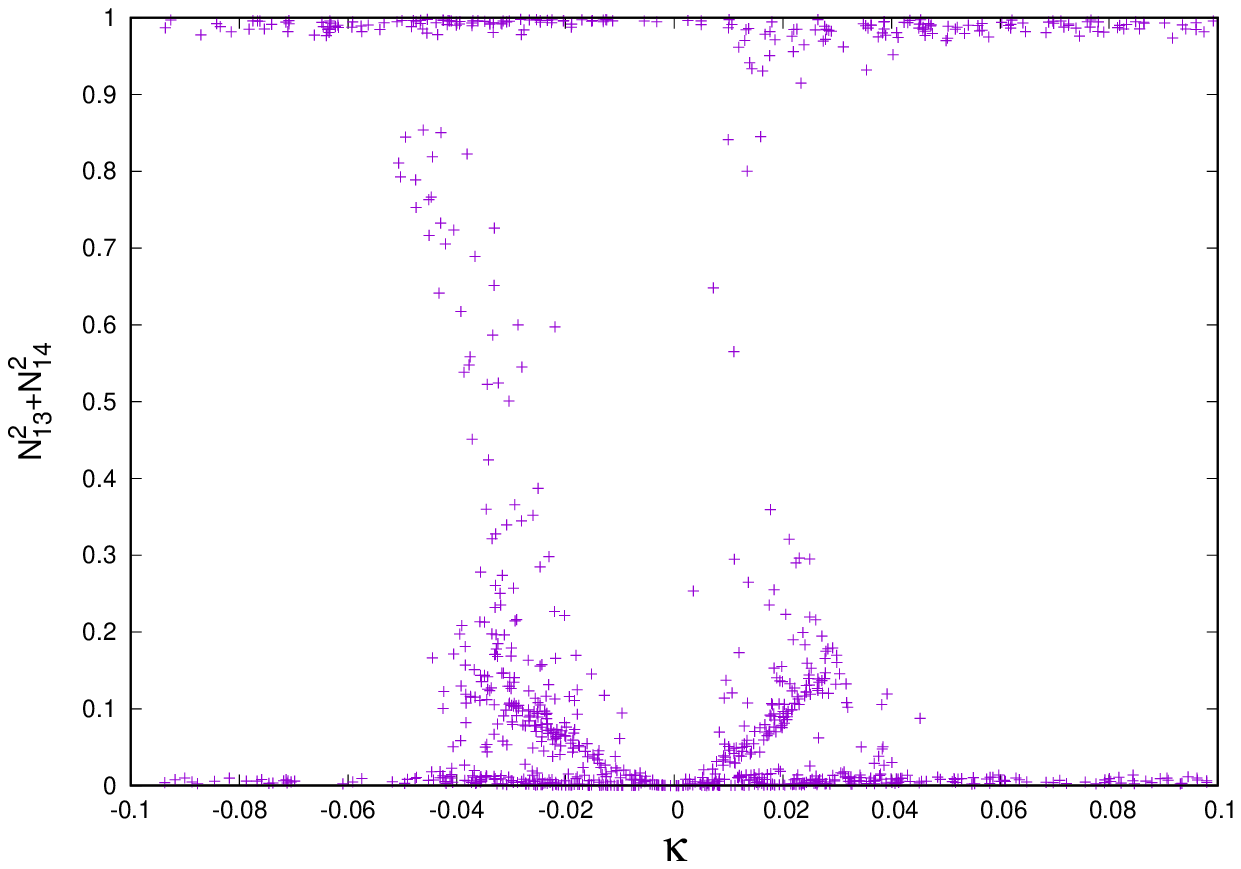} &\includegraphics[scale=0.33]{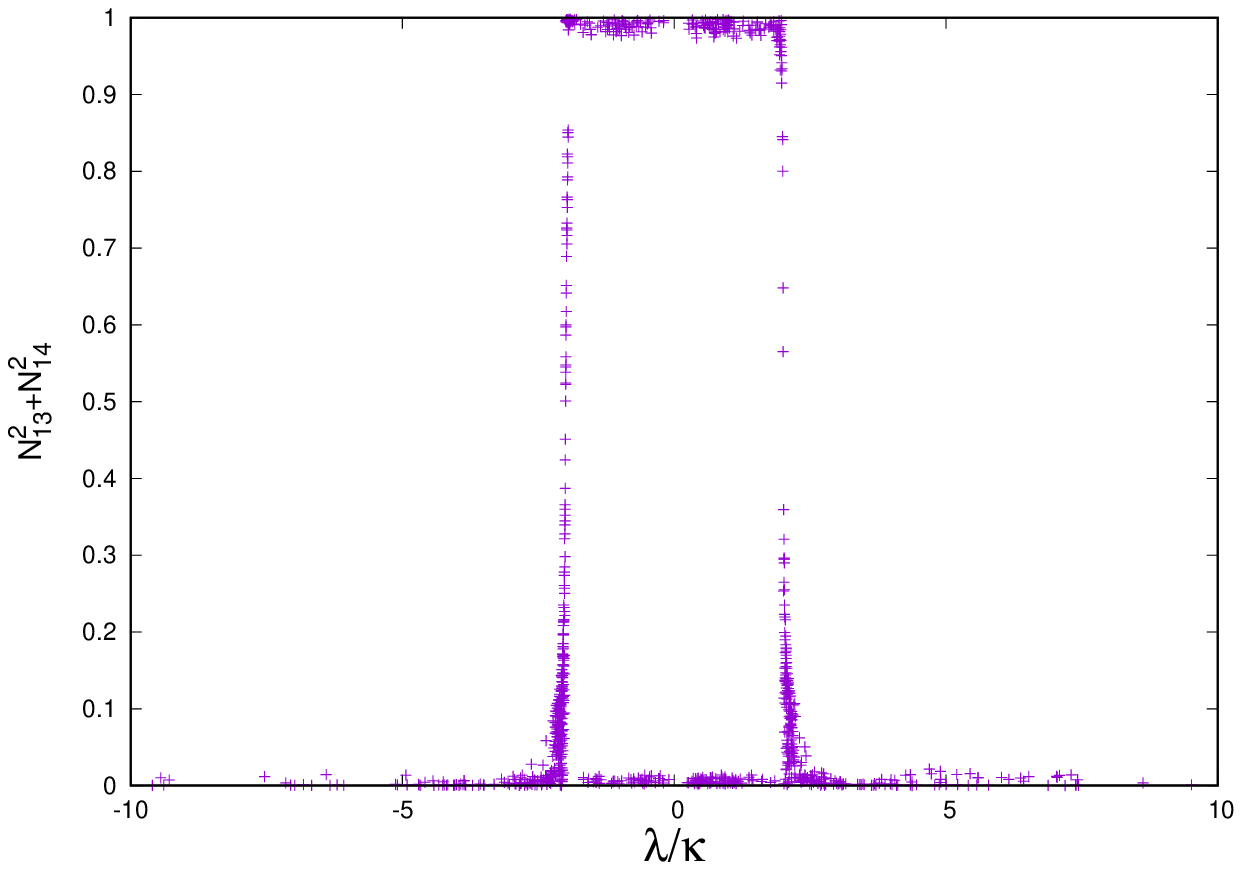}&\includegraphics[scale=0.33]{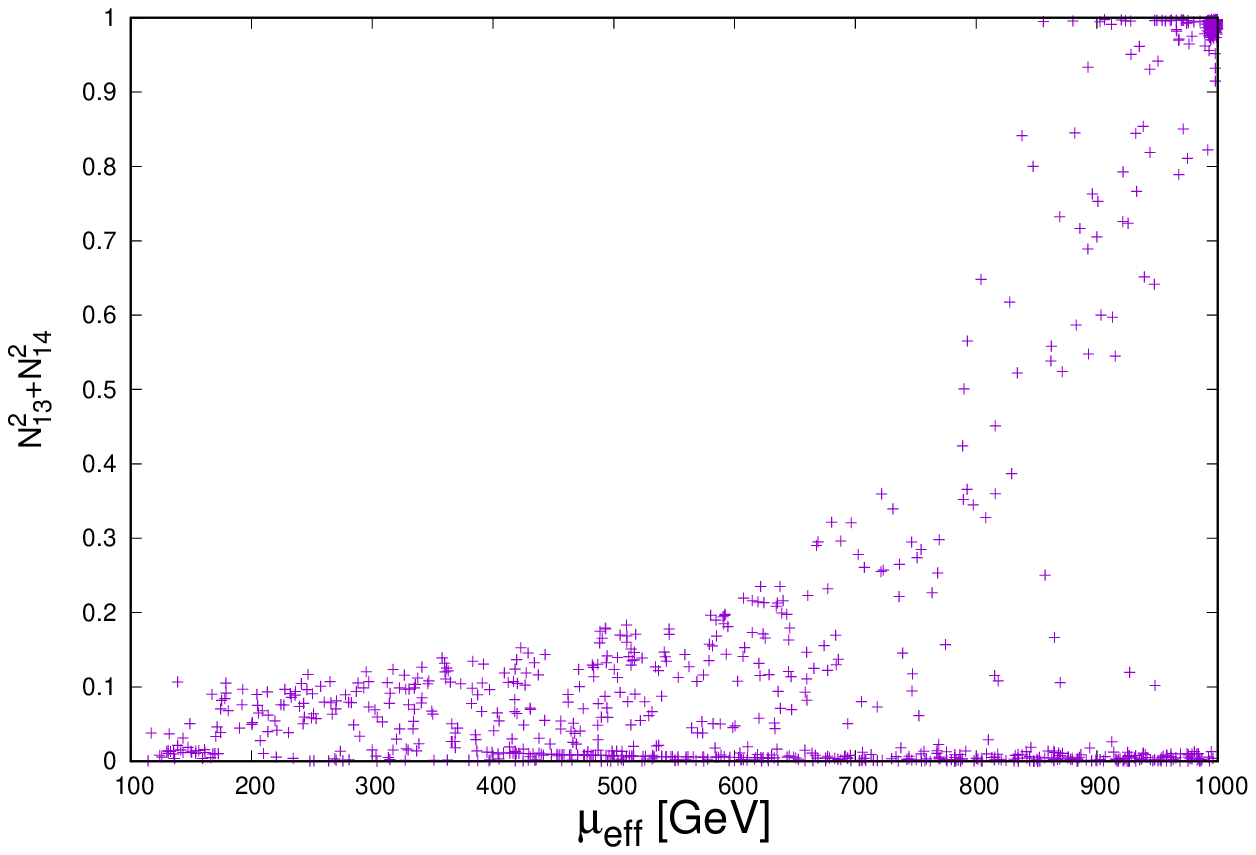}\\
 \includegraphics[scale=0.33]{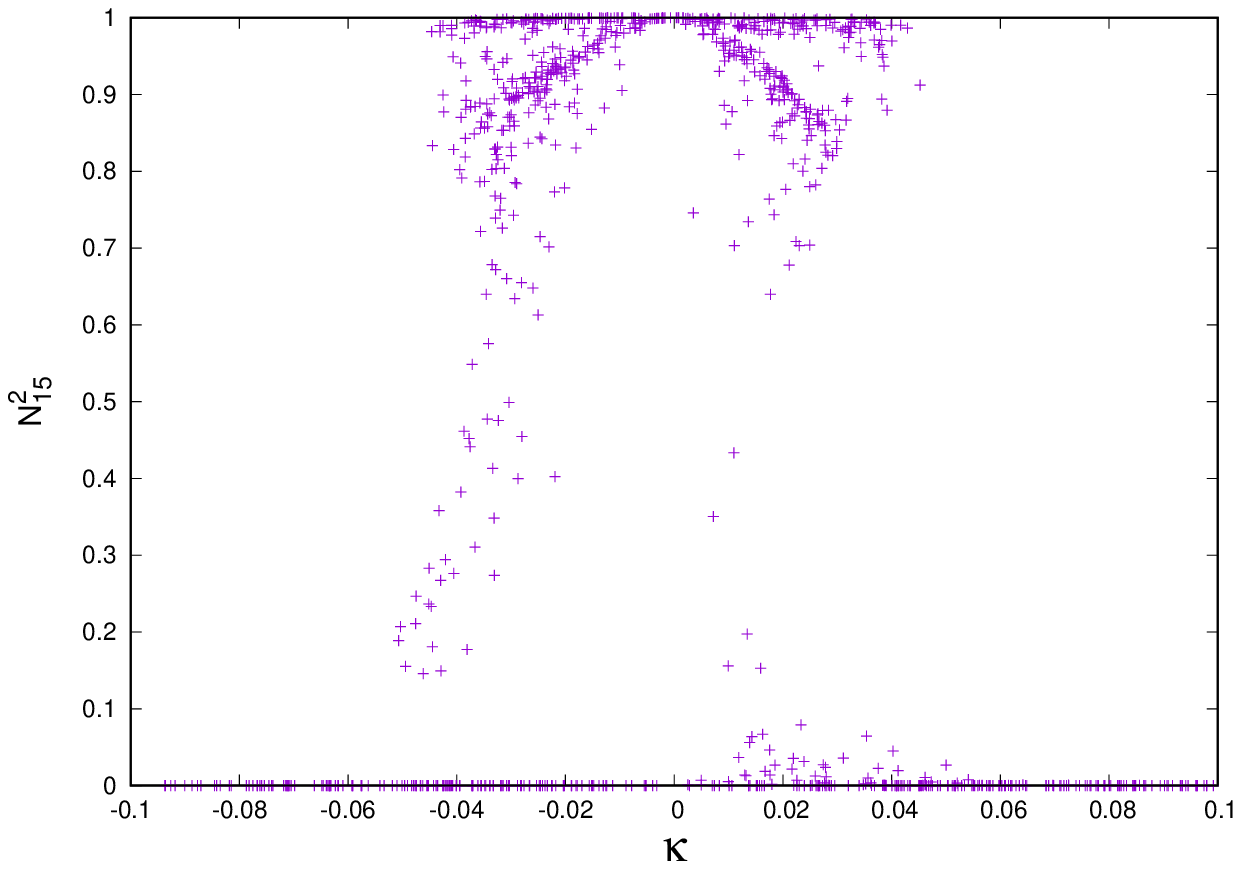} & \includegraphics[scale=0.33]{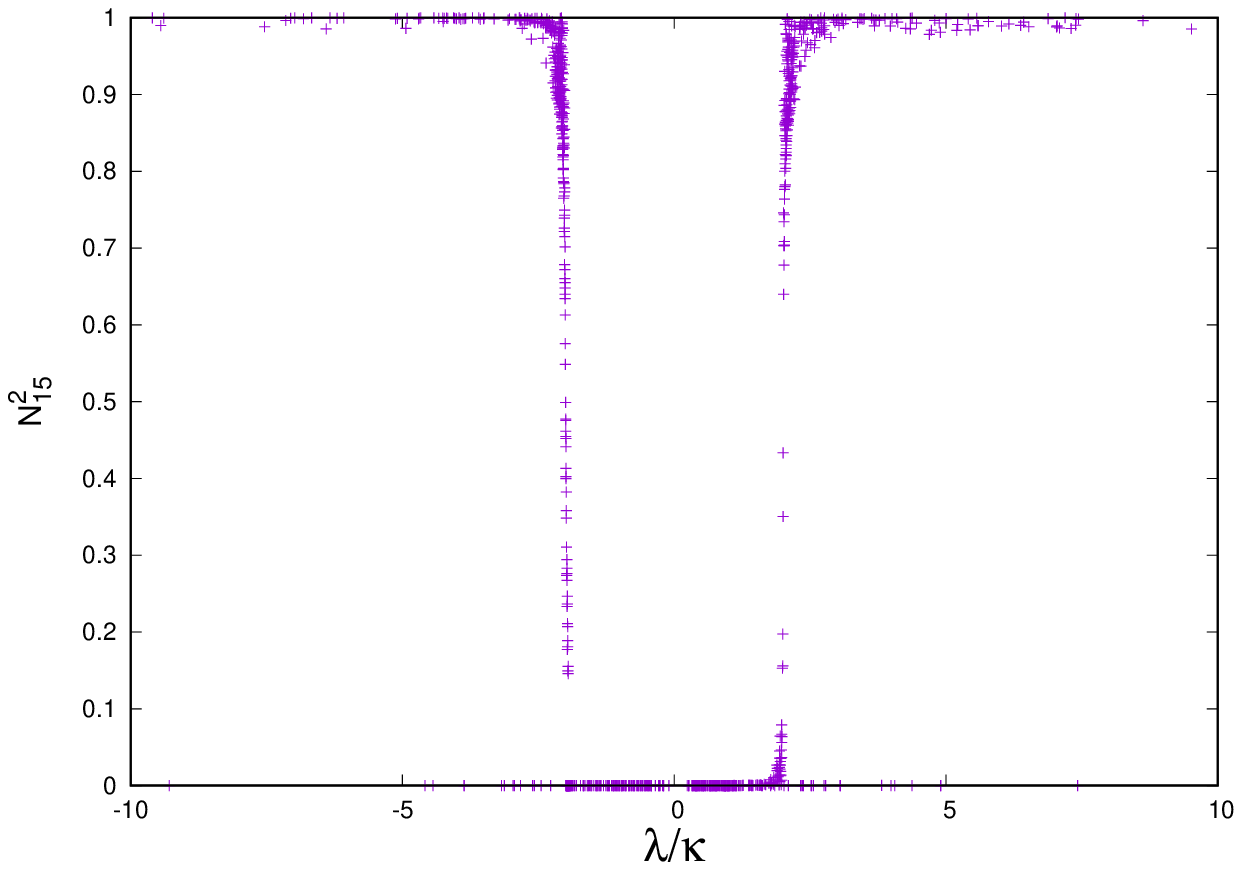}&\includegraphics[scale=0.33]{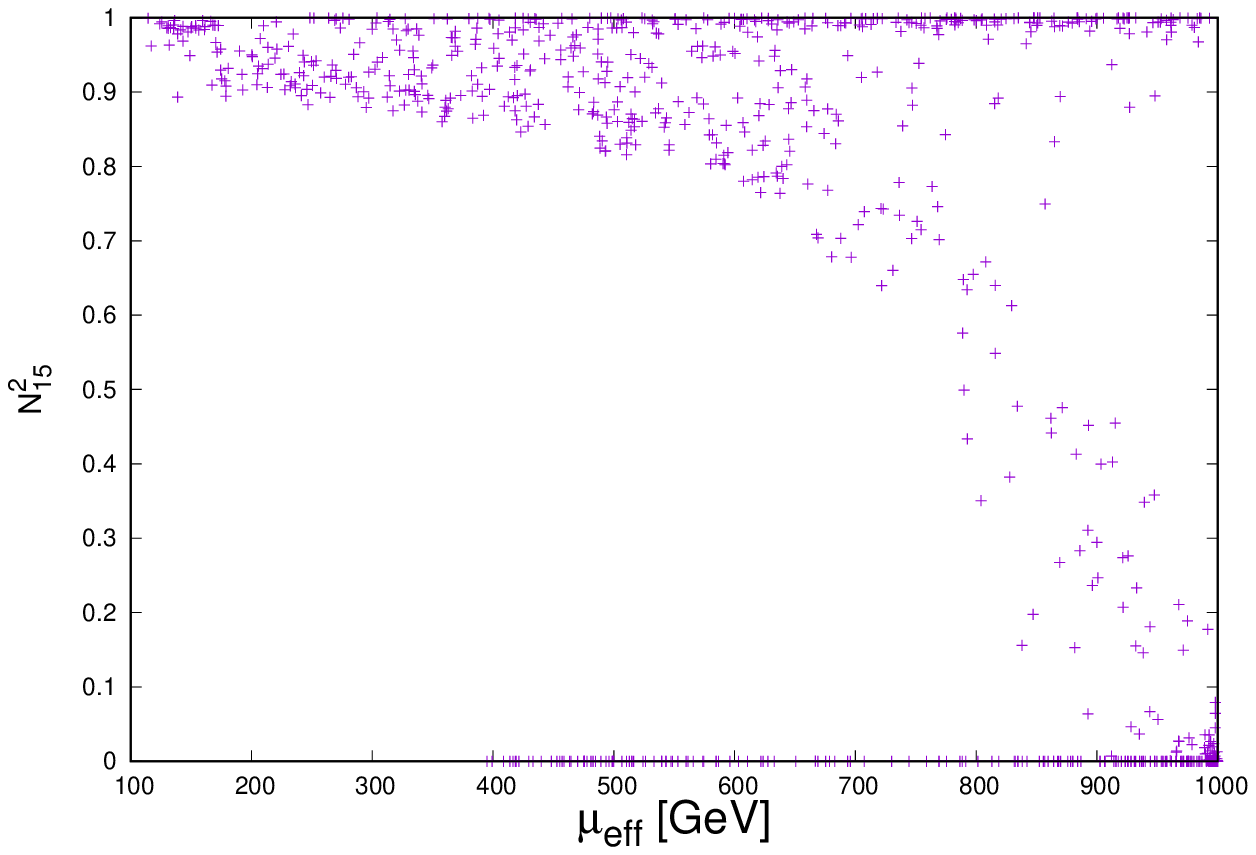}
 \end{tabular}

\caption{The mass components of the lightest neutralino, $\tilde{\chi}^0_1$, versus $\kappa$ (left-panels), $\lambda/\kappa$ (middle-panels) and $\mu_{\rm eff}$ (right-panels).}
  \label{fig4}
\end{figure}

\begin{figure}
 \centering\begin{tabular}{cc}
 \includegraphics[scale=0.5]{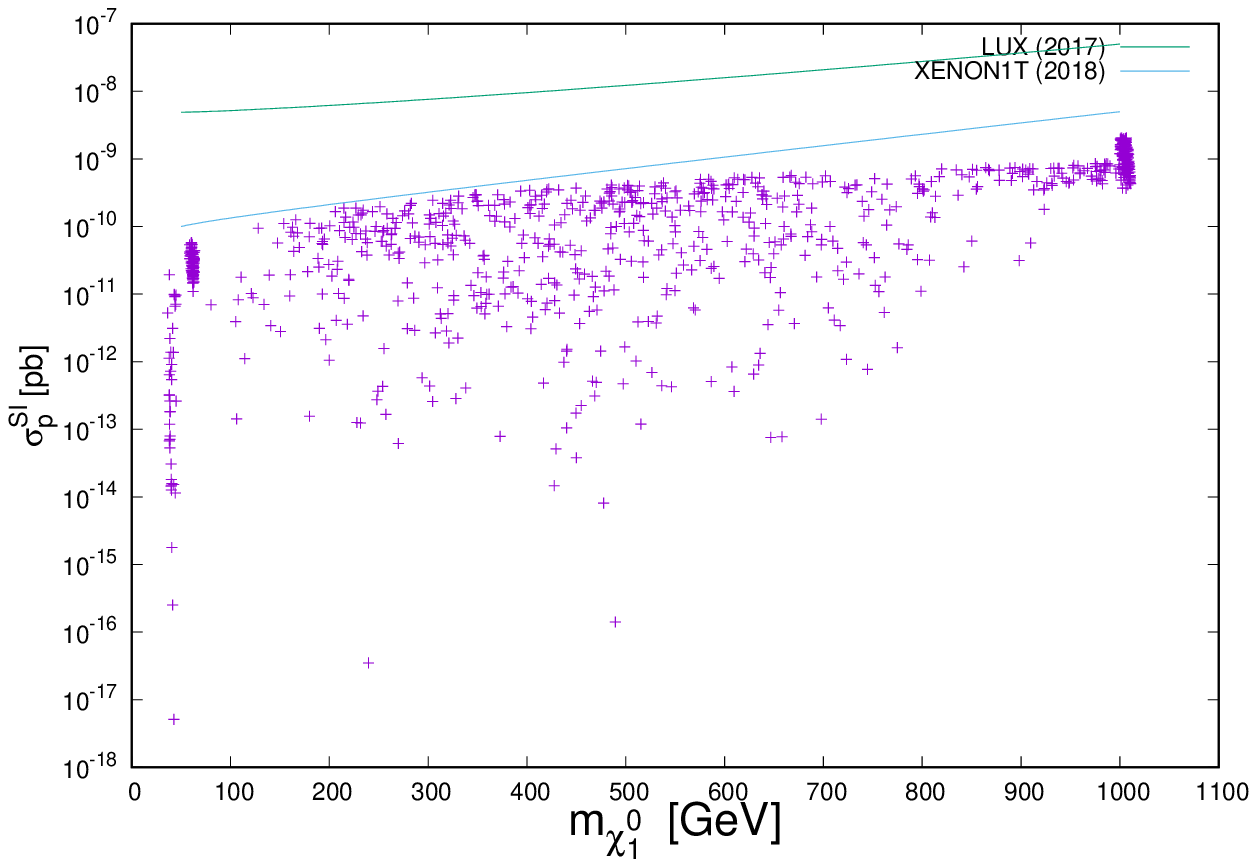} &\includegraphics[scale=0.5]{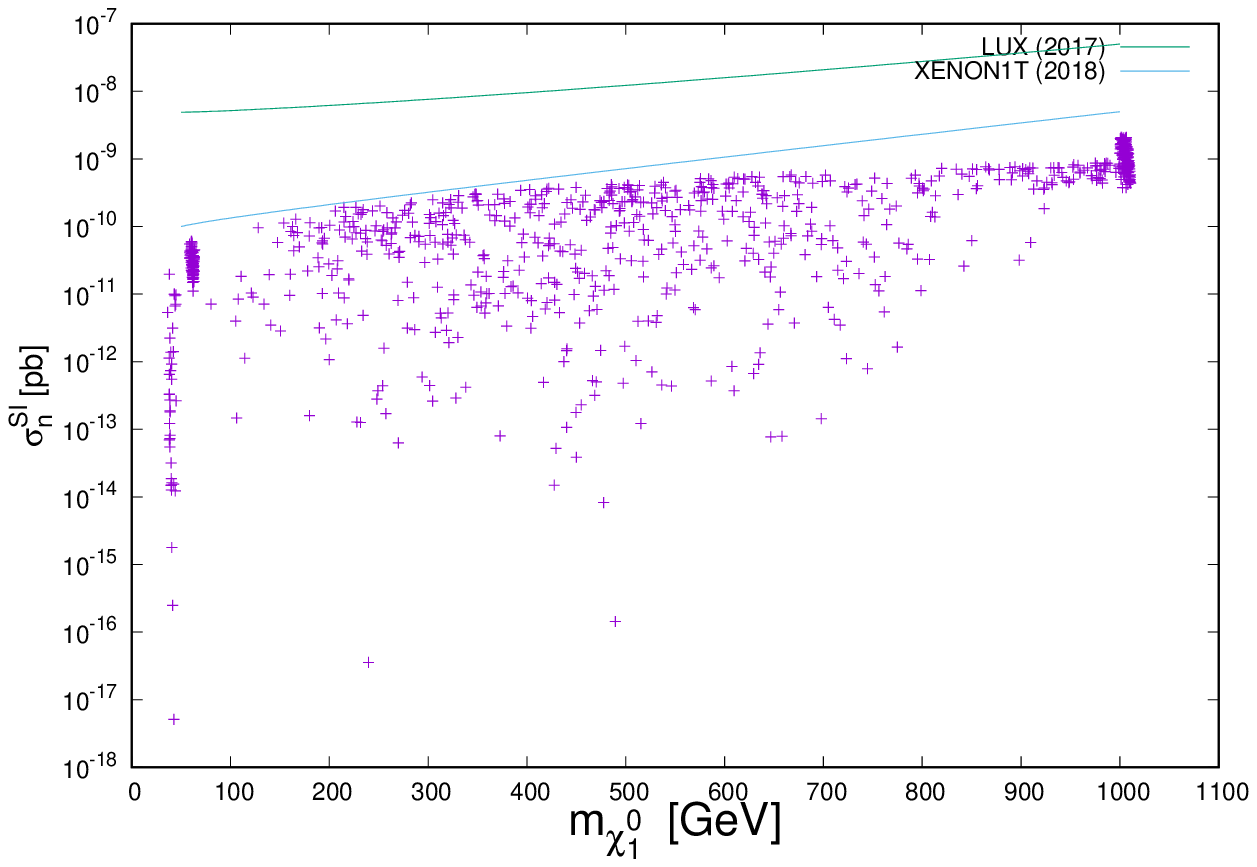}\\
  \includegraphics[scale=0.5]{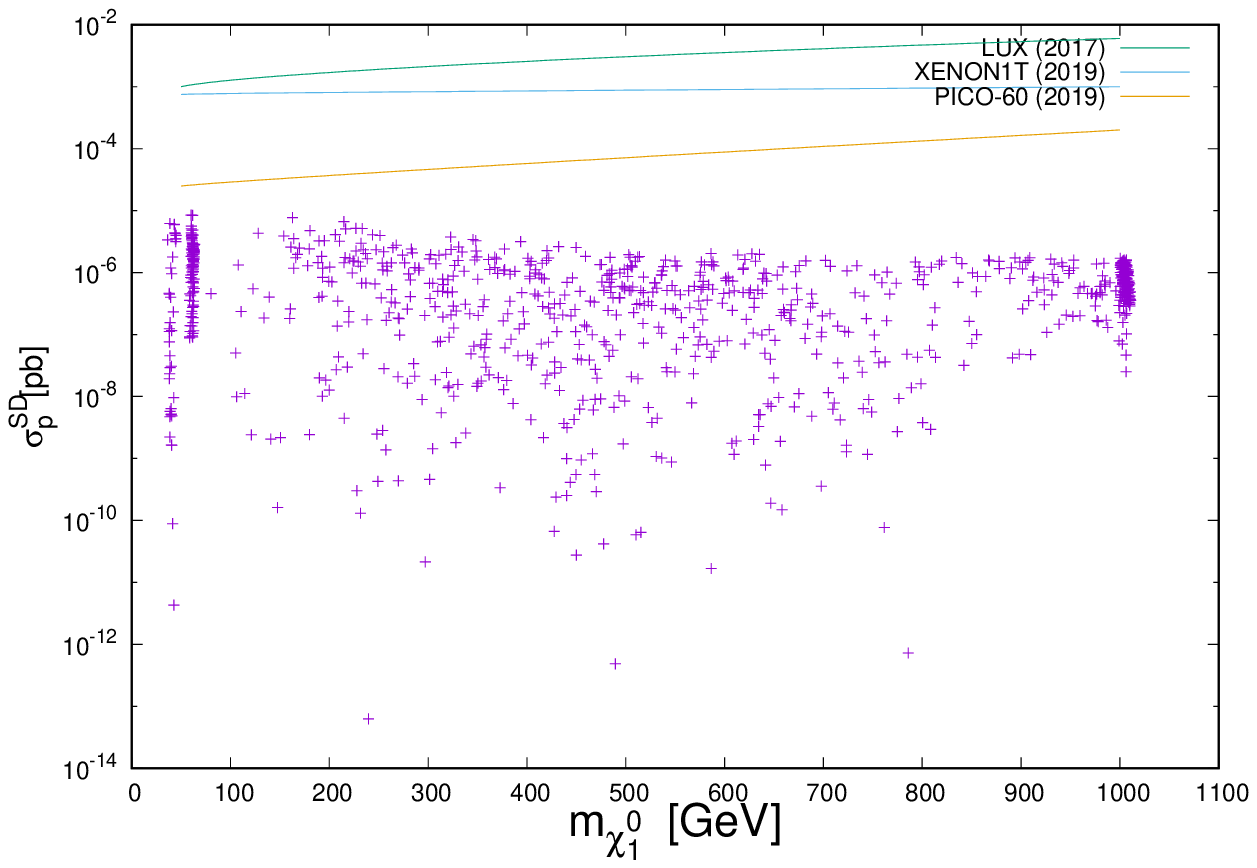} &\includegraphics[scale=0.5]{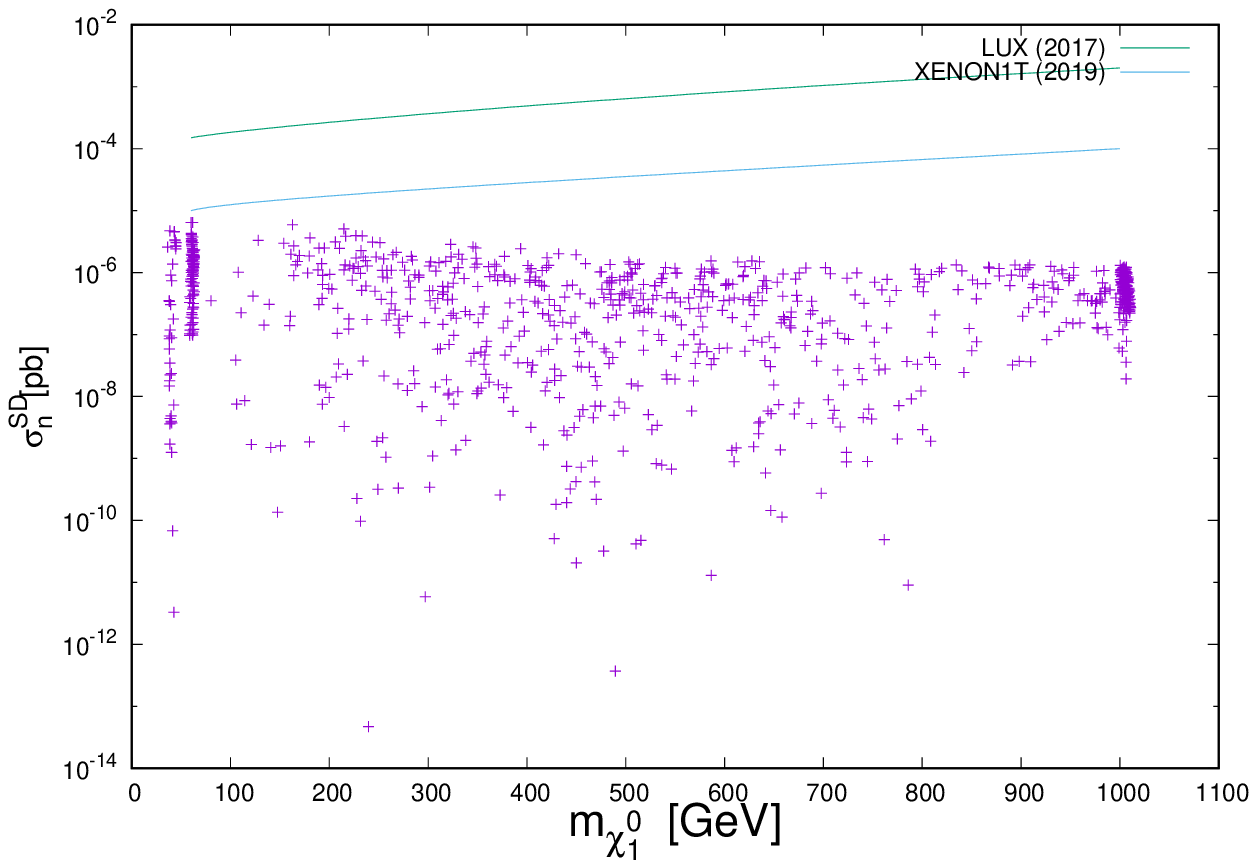}\\

 \end{tabular}

\caption{Spin-independent (SI) and spin-dependent (SD)
cross sections of DM-proton scattering and of DM-neutron scattering.
The solid lines correspond to exclusions on $\sigma^{SI}$ and $\sigma^{SD}$ from LUX, XENON-1T and PICO-60  \cite{LUX:2017ree,XENON:2018voc,XENON:2019rxp,PICO:2019vsc}.}
\label{fig5}
\end{figure}

\begin{figure}
 \centering\begin{tabular}{cc}
 \includegraphics[scale=0.5]{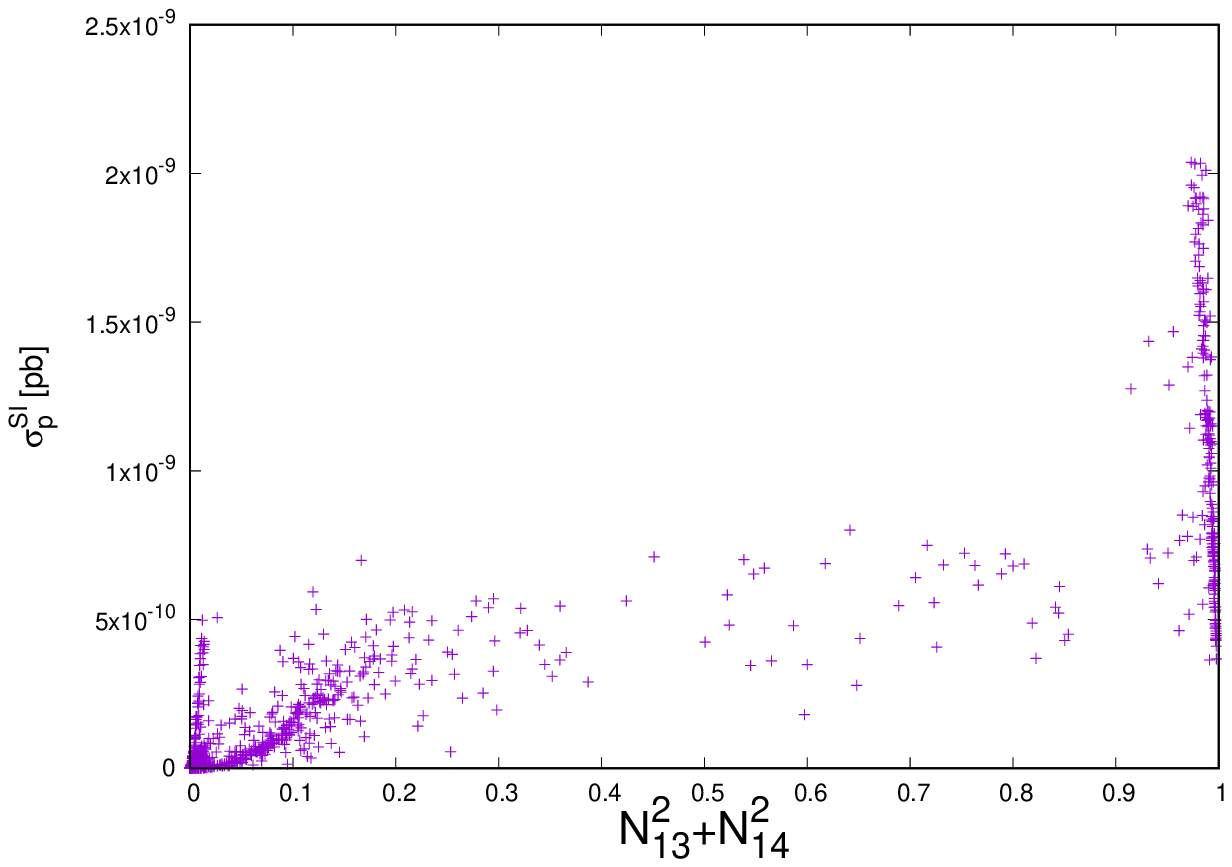} &\includegraphics[scale=0.5]{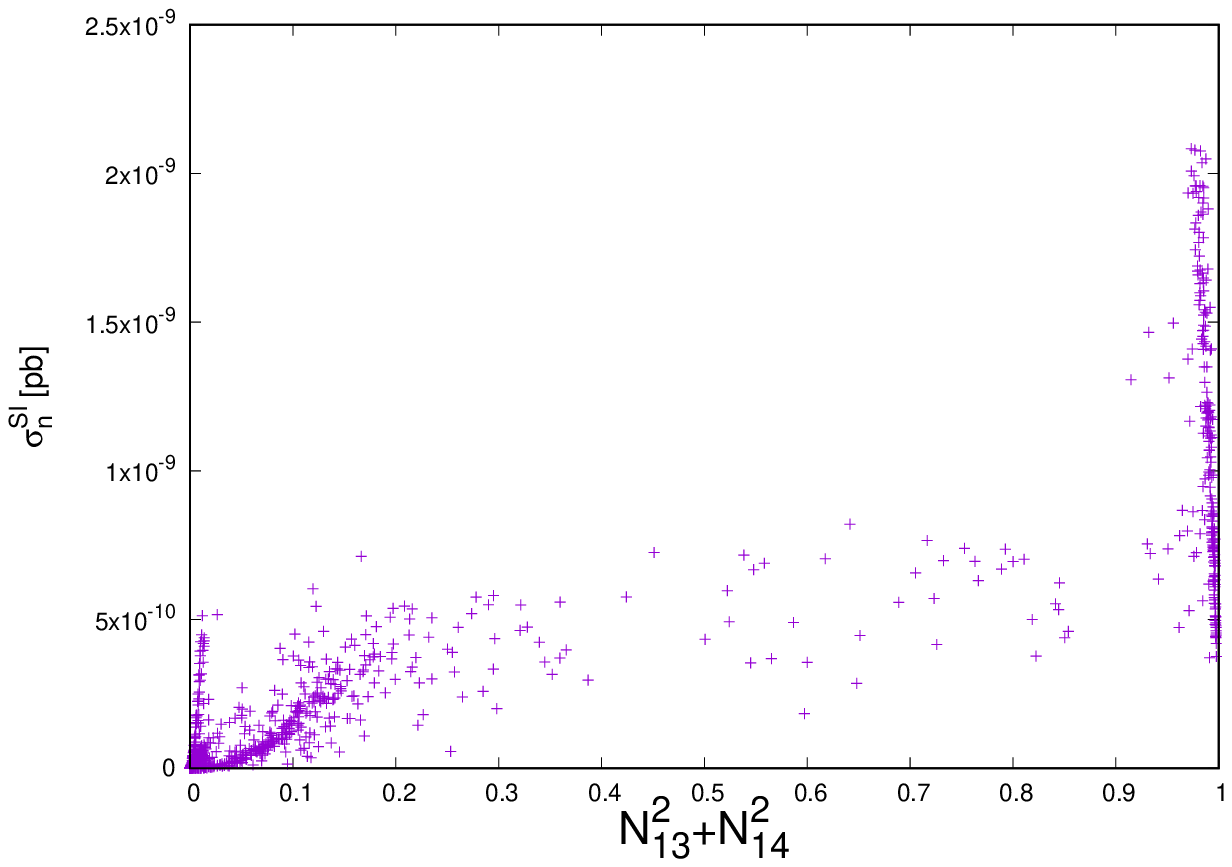}\\
  \includegraphics[scale=0.5]{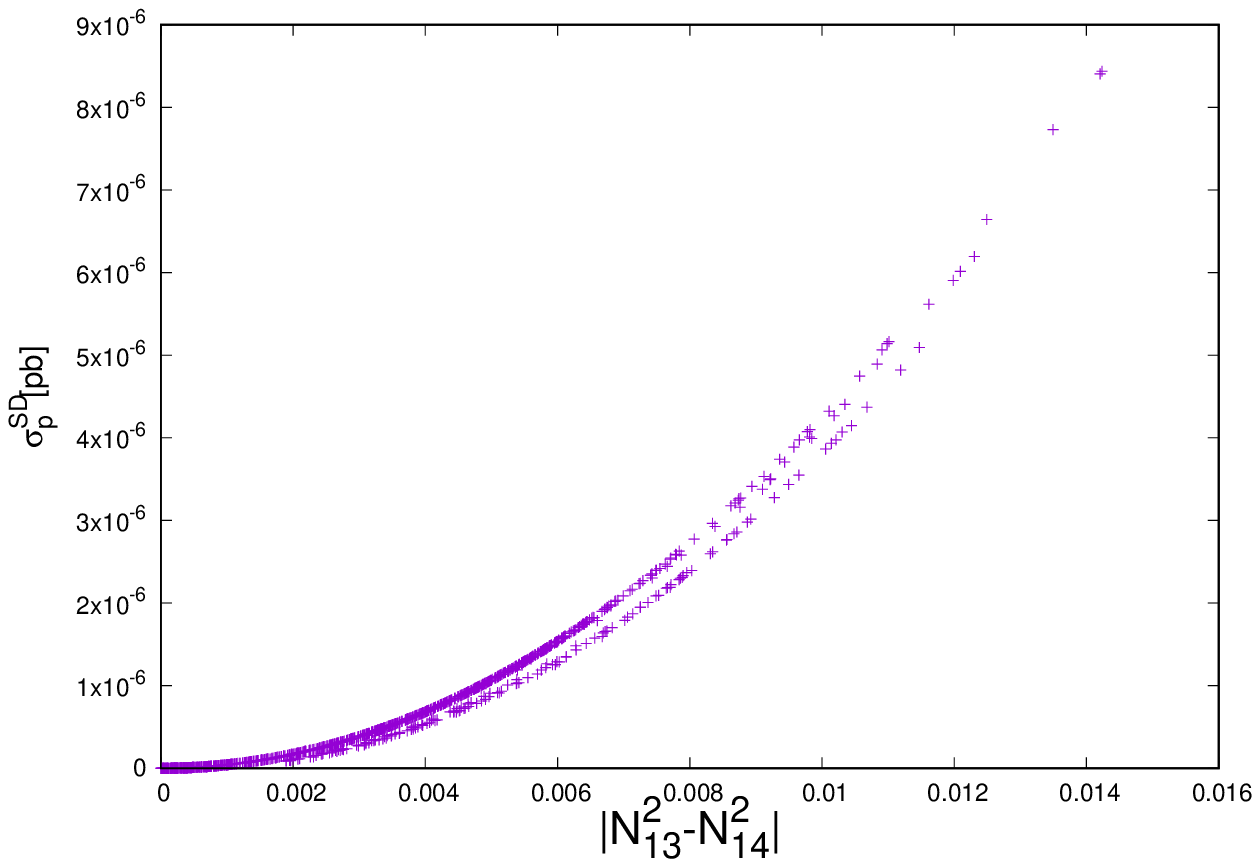} &\includegraphics[scale=0.5]{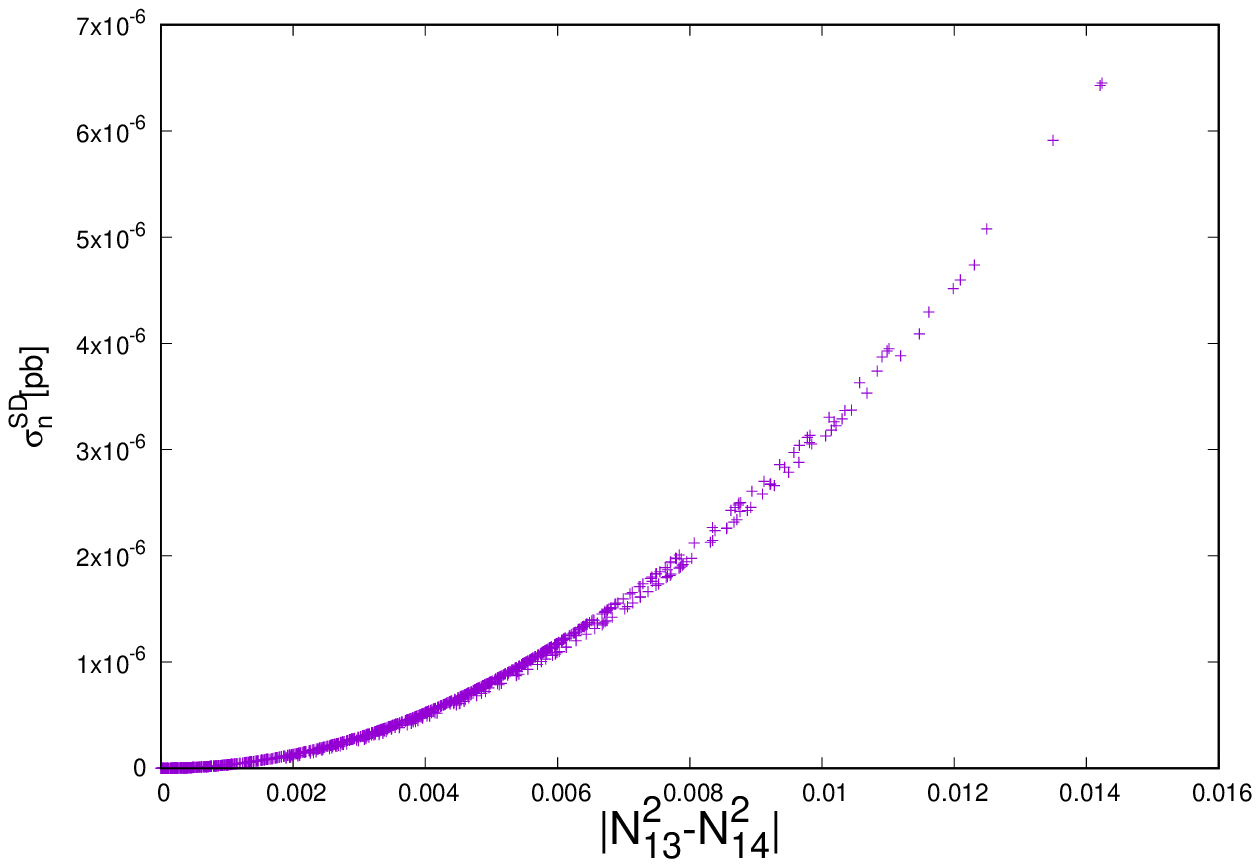}\\

 \end{tabular}

\caption{The higgsino component of the $\tilde{\chi}^0_1$ versus both $\sigma^{SI}_p$ and $\sigma^{SI}_n$ (upper-panels) and 
 the difference of the higgsino components of the $\tilde{\chi}^0_1$ versus both $\sigma^{SD}_p$ and $\sigma^{SD}_n$ (bottom-panels).}
\label{fig6}
\end{figure}

\begin{figure}
 \centering\begin{tabular}{cc}
 \includegraphics[scale=0.5]{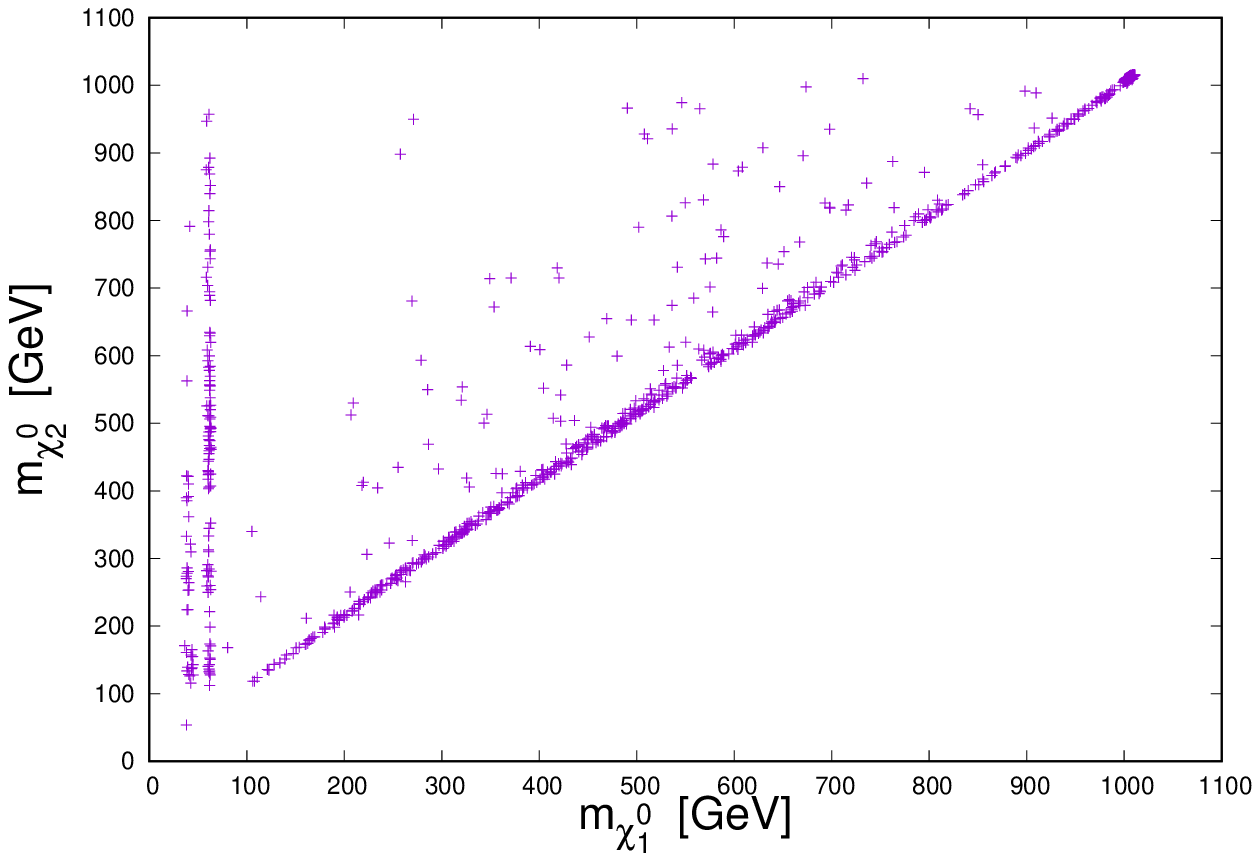} &\includegraphics[scale=0.5]{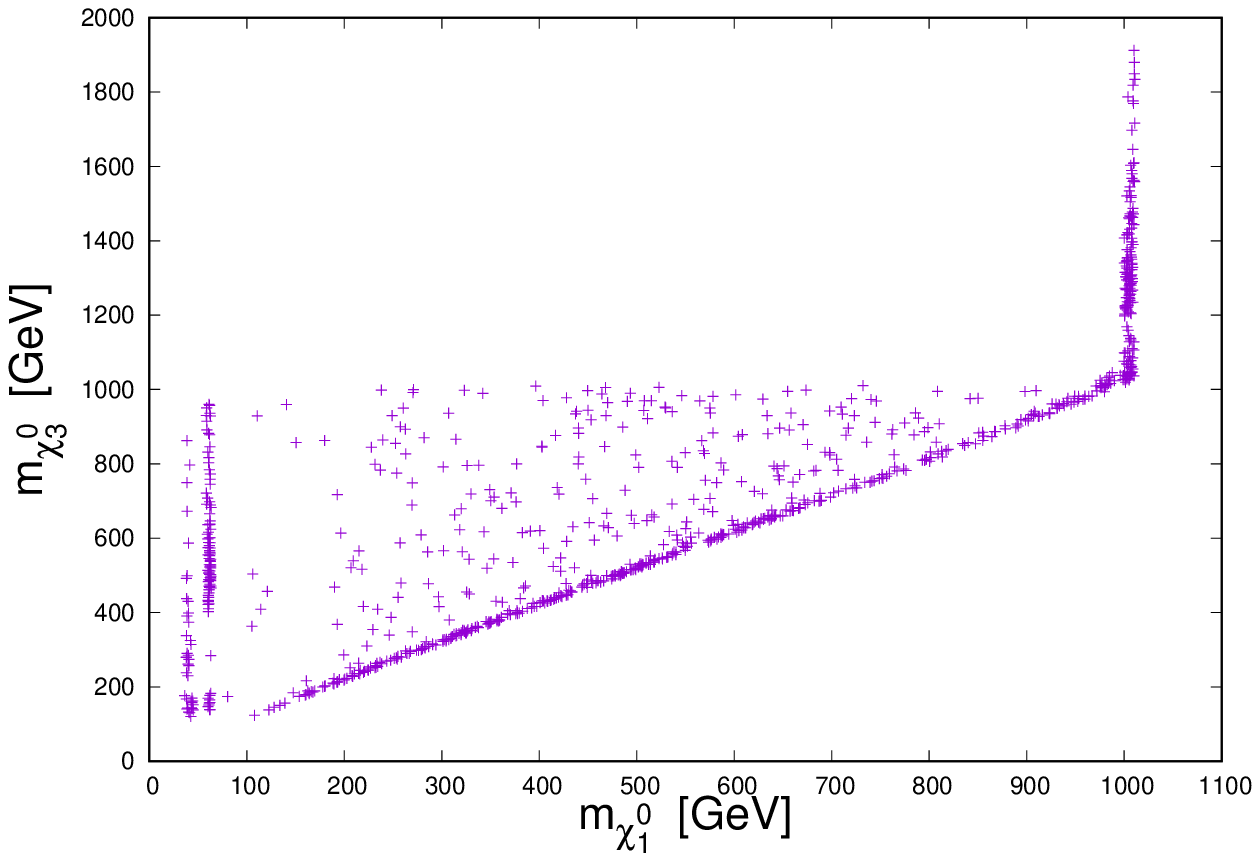}\\
   \end{tabular}
 \centering
  \includegraphics[scale=0.5]{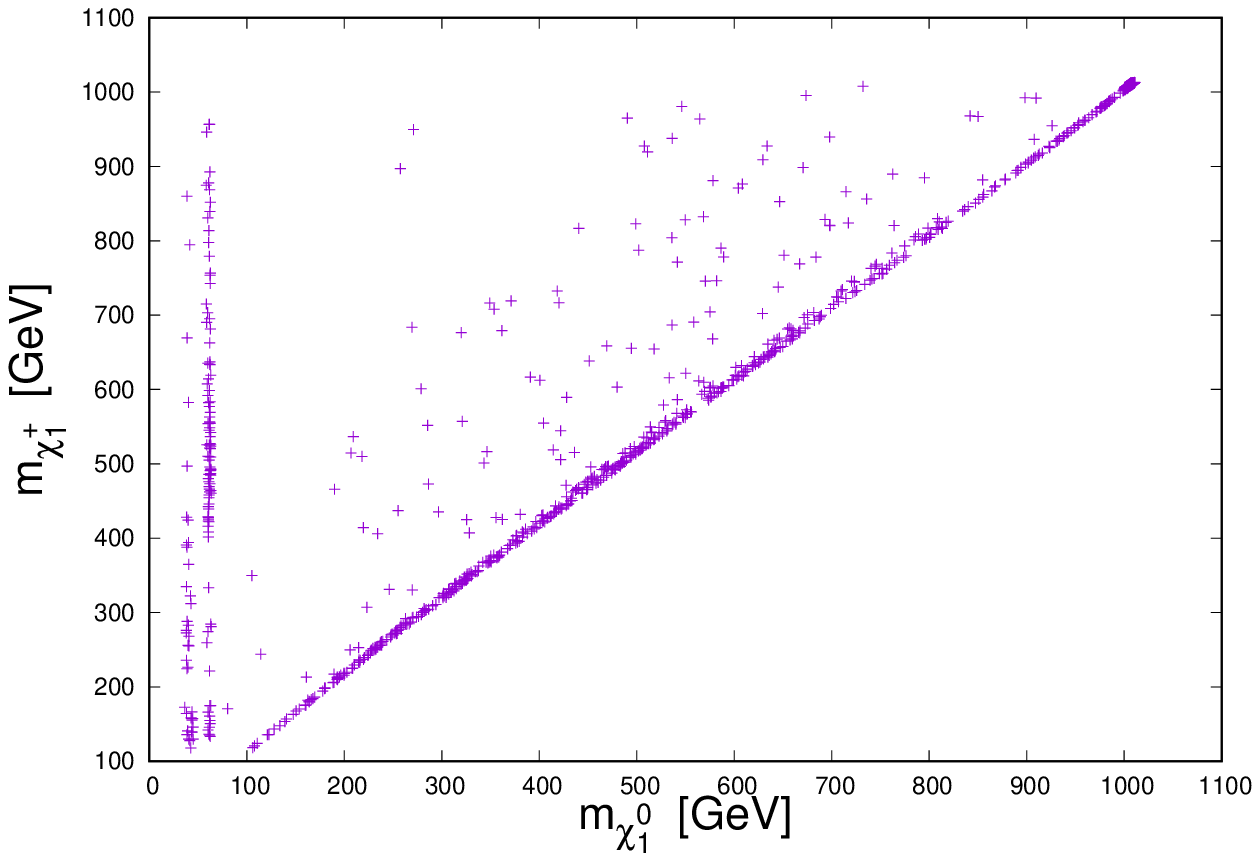}

\caption{The lightest neutralino mass, $m_{\tilde{\chi}^0_1}$,
versus the second-lightest neutralino mass, $m_{\tilde{\chi}^0_2}$, third neutralino mass, $m_{\tilde{\chi}^0_3}$ and  
lightest chargino mass, $m_{\tilde{\chi}^{\pm}_1}$.}

\label{fig7}
\end{figure}

\begin{figure}
 \centering\begin{tabular}{cc}
 \includegraphics[scale=0.5]{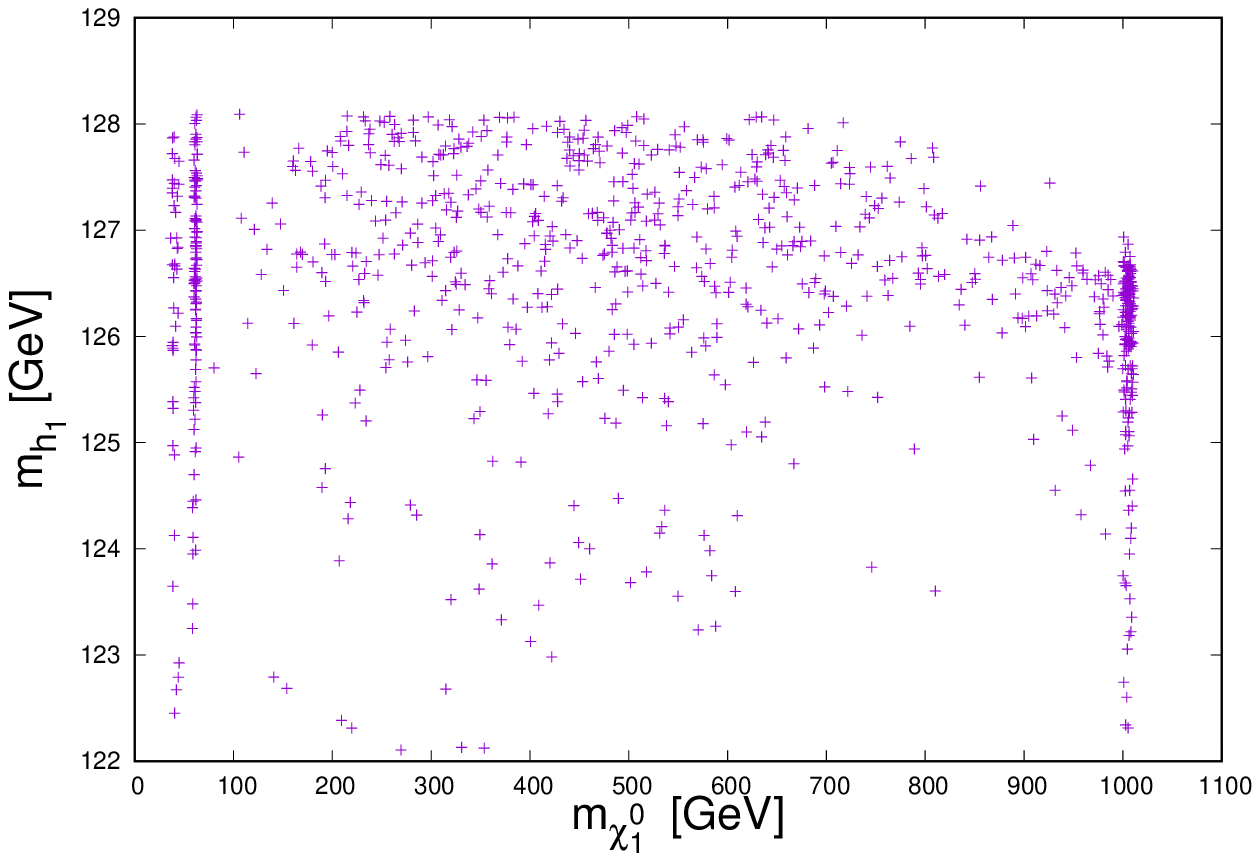} &\includegraphics[scale=0.5]{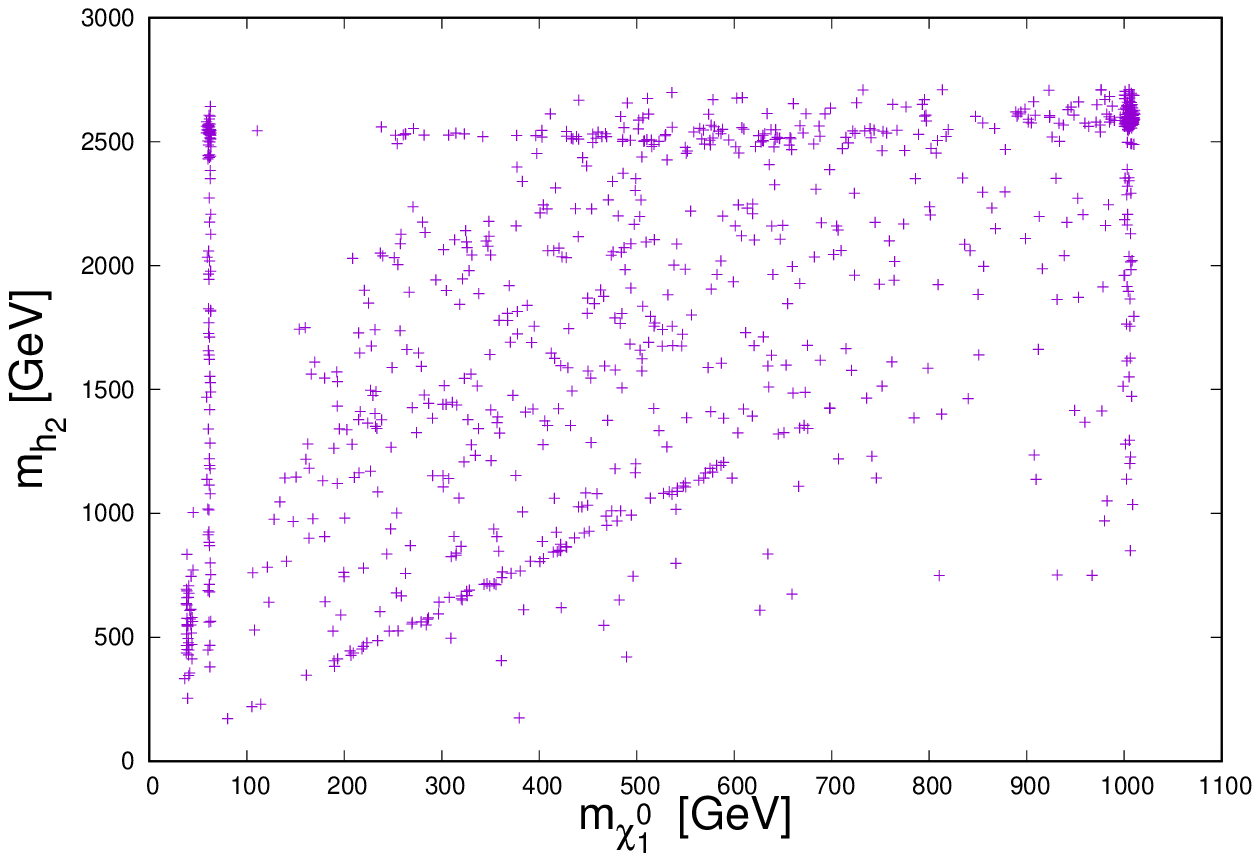}\\
   \end{tabular}
 \centering
  \includegraphics[scale=0.5]{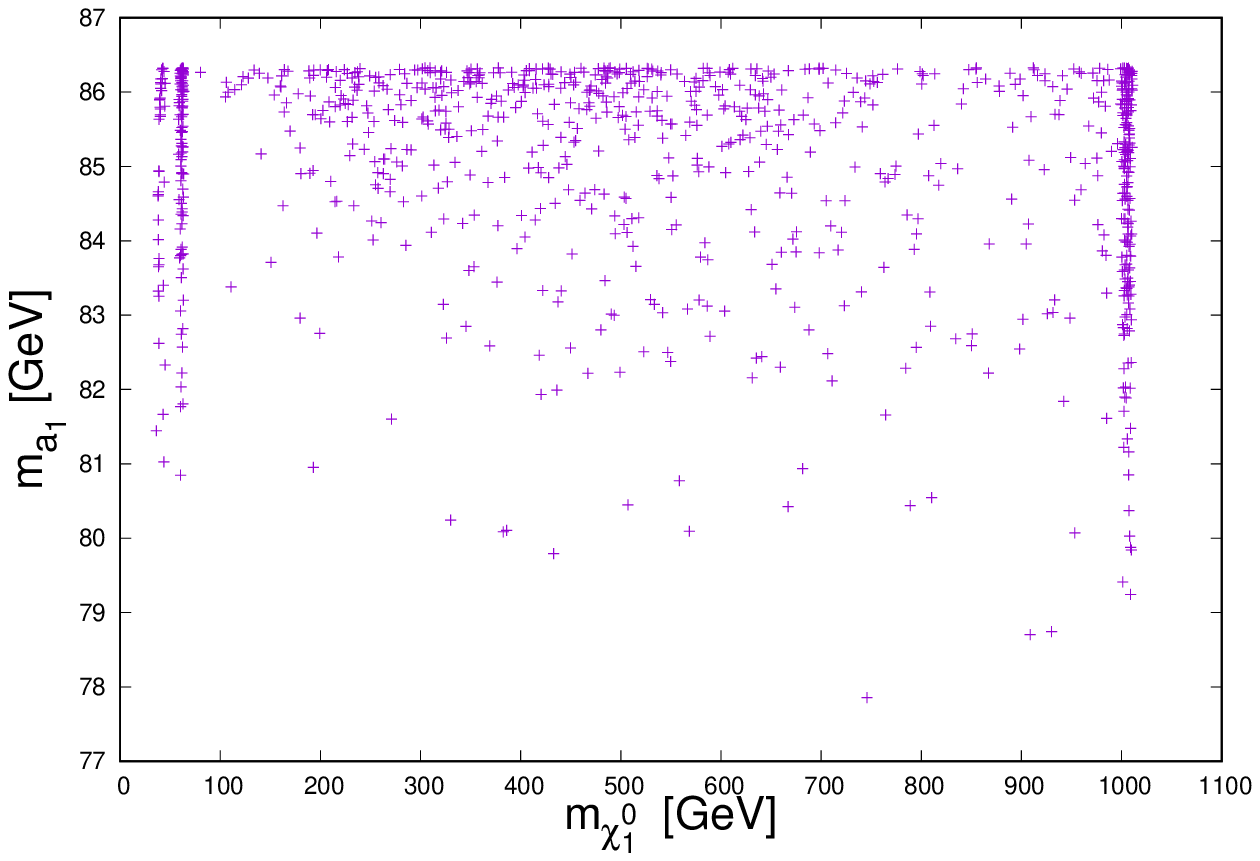}

\caption{The lightest neutralino mass, $m_{\tilde{\chi}^0_1}$,
versus the lightest and second-lightest neutral scalar Higgs boson masses, $m_{h_1}$ and $m_{h_2}$, and the lightest pseudo-scalar
neutral Higgs mass, $m_{a_1}$.}
\label{fig8}
\end{figure}

\begin{figure}
 \centering\begin{tabular}{c}
 \includegraphics[scale=0.6]{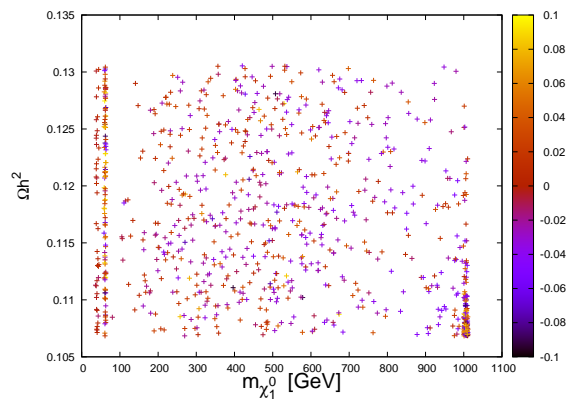} 
 
 \end{tabular}

\caption{The relic density, $\Omega h^2$, versus the lightest neutralino mass $m_{\tilde{\chi}^0_1}$. The scale on the right represents
the values of $\kappa$ in the figure.}
\label{fig9}
\end{figure}

\newpage
\label{sect:summa}
\section{Conclusions}
In the NMSSM framework, the $\mu$-problem of the MSSM superpotential is solved by extending the MSSM with one more neutral singlet superfield.
This introduction of the singlet superfield makes the Higgs and neutralino sectors of the NMSSM are phenomenologically richer than that of the MSSM,
giving rise to seven Higgs and five neutralino states compared to only five Higgses and four neutralinos in the MSSM. 
In this work, we targeted a specific scenario carrying out small ranges of $\lambda$ and $\kappa$ in order to 
 examine the lightest neutralino features as a DM candidate. For this to be achieved, a random scan across the NMSSM parameter space
 has been performed, taking into consideration theoretical and experimental constraints, including the relic density
 in addition to direct and indirect detection limits.
 Out of the input parameters, the analysis of the surviving points has shown that $\kappa$ and $\mu_{\rm eff}$
 in addition to the ratio $\lambda/\kappa$ are the major contributors
 to the mass and properties of the lightest neutralino.
 These contributors have revealed an additional potential dominating component of the lightest neutralino,
 which is the singlino ($\tilde{S}^0$). We have found that
 the LSP dark matter is a singlino-dominated neutralino for most of the sampled points for the chosen scenario. 
 Thus, contrary to expectations, the NMSSM is not comparable to the MSSM as $\lambda$ and $\kappa$ are very small, $\lesssim 0.1$.
  If the LSP is singlino-dominated, it may hardly couple to any SM particle. In this case, the non-observation of WIMP scattering
  could make the lightest neutralino of the NMSSM an excellent candidate for the DM.
 We also found that the LSP dark matter can be either a higgsino-dominated for $ m_{\tilde{\chi}^0_1} \gtrsim$ 850 GeV or a bino-dominated for $ m_{\tilde{\chi}^0_1} \lesssim$
 725 GeV in some area of the NMSSM parameter space.
 
 Furthermore, we have computed the SI and SD cross sections for elastic WIMP-proton scattering and elastic WIMP-neutron scattering 
 and found that $\sigma^{SI}_p \approx  1.02$ $\sigma^{SI}_n$ and $\sigma^{SD}_p \approx  0.76$ $\sigma^{SD}_n$.
The SI cross section and DM annihilation depend on the masses of Higgs bosons.
 The lightest neutral scalar Higgs boson mass $m_{h_{1}}$ has been found to be the SM-like Higgs in our parameter space,
 while the second-lightest neutral scalar
 Higgs boson $m_{h_{2}}$ has been found to have a mass up to 2500 GeV. The lightest pseudo-scalar neutral Higgs $m_{a_{1}}$ is always highly singlet with a mass smaller than
 $m_Z$. The other heavy Higgses typically have the same masses.
  
Finally, we have found that the NMSSM can allow the predicted $m_{\tilde{\chi}^0_1}$ ranges from about 60 GeV up to about 1 TeV.
The correct relic density can be fulfilled for different values of $m_{\tilde{\chi}^0_1}$ and for any neutralino composition
whether it is a singlino-dominated, a higgsino-dominated, a bino-dominated, or a mixture of them. \\

\end{document}